\documentclass[10pt]{article}  
\usepackage[margin=1in]{geometry}
\usepackage{url}
\usepackage[utf8]{inputenc}
\usepackage[english]{babel}
\usepackage{amsthm}

\usepackage[mathscr]{euscript} % For probability P symbols
\let\euscr\mathscr \let\mathscr\relax % just so we euscr
\usepackage{cite,mathtools} % to group citations
\usepackage{latexsym}
\usepackage{amssymb,amsbsy,amsmath,amsfonts,amssymb,amscd,amsfonts,mathrsfs}
\usepackage{graphicx}
\usepackage{color} 

\usepackage{txfonts}
\usepackage{mathtools}
\DeclareMathOperator*{\argmax}{arg\,max}
\DeclareMathOperator*{\supp}{Supp}

\newtheorem{remark}{Remark}[]

\numberwithin{equation}{section}

\usepackage{titling}
\predate{\begin{flushleft}}
\postdate{\end{flushleft}}
\allowdisplaybreaks
\usepackage[font=rm, labelfont=bf]{caption}

\title{Individual-based and continuum models of phenotypically heterogeneous growing cell populations}

\author{\normalsize
 Fiona R Macfarlane$^{1,*}$,
 Xinran Ruan$^{2}$,
 Tommaso Lorenzi$^{3,*}$
}

\date{ \small \hspace{3em}  $^{1}$ School of Mathematics and Statistics, University of St Andrews, United Kingdom;\\
 \hspace{3em}  $^{2}$ School of Mathematical Sciences, Capital Normal University, 100048 Beijing, P.R. China;\\
 \hspace{3em}  $^{3}$ Department of Mathematical Sciences ``G. L. Lagrange'', Politecnico di Torino, 10129 Torino, Italy;\\
 \ \\
\hspace{3.5em}\footnotesize{$^*$Corresponding authors: frm3@st-andrews.ac.uk; tommaso.lorenzi@polito.it}}
\begin{document}
\maketitle

\begin{abstract}
Existing comparative studies between individual-based models of growing cell populations and their continuum counterparts have mainly been focused on homogeneous populations, in which all cells have the same phenotypic characteristics. However, significant intercellular phenotypic variability is commonly observed in cellular systems. In light of these considerations, we develop here an individual-based model for the growth of phenotypically heterogeneous cell populations. In this model, the phenotypic state of each cell is described by a structuring variable that captures intercellular variability in cell proliferation and migration rates. The model tracks the spatial evolutionary dynamics of single cells, which undergo pressure-dependent proliferation, heritable phenotypic changes and directional movement in response to pressure differentials. We formally show that the continuum limit of this model comprises a non-local partial differential equation for the cell population density function, which generalises earlier models of growing cell populations. We report on the results of numerical simulations of the individual-based model which illustrate how proliferation-migration tradeoffs shaping the evolutionary dynamics of single cells can lead to the formation, at the population level, of travelling waves whereby highly-mobile cells locally dominate at the invasive front, while more-proliferative cells are found at the rear. Moreover, we demonstrate that there is an excellent quantitative agreement between these results and the results of numerical simulations and formal travelling-wave analysis of the continuum model, when sufficiently large cell numbers are considered. We also provide numerical evidence of scenarios in which the predictions of the two models may differ due to demographic stochasticity, which cannot be captured by the continuum model. This indicates the importance of integrating individual-based and continuum approaches when modelling the growth of phenotypically heterogeneous cell populations.
\end{abstract}

%%%%%%%%%%%%%%%%%%%%%%%%%%%%%%%%%%%%%%%%%
%% Introduction %% 
%%%%%%%%%%%%%%%%%%%%%%%%%%%%%%%%%%%%%%%%%
\section{Introduction}
\label{sec:introduction}
Deterministic continuum models for the growth of cell populations have been increasingly used as theoretical tools to support empirical research regarding a broad spectrum of aspects of the development of solid tumours and living tissues. {{These models comprise partial differential equations (PDEs) that describe the evolution of cellular densities (or cell volume fractions) in response to pressure gradients that are generated by population growth, which can be mechanically-regulated~\cite{chaplain2019derivation,ranft2010fluidization}, nutrient-limited~\cite{Giverso_avascular}, pressure-dependent~\cite{bubba2020hele,david2022asymptotic,lorenzi2017interfaces} or regulated by a combination of these mechanisms~\cite{bresch2010computational,ciarletta2011radial,Gallinato,perthame2014some,lowengrub2009nonlinear}.}}
These models are amenable not only to numerical simulations but also to analytical approaches, which enable a complete exploration of the model parameter space. This permits a precise identification of the validity domain of the results obtained and ensures higher robustness and precision of the conclusions drawn therefrom, which ultimately provides a more in-depth theoretical understanding of the underlying cellular dynamics~\cite{kuznetsov2021improving,lorenzi2022cancer}. 

Ideally, instead of defining such PDE models on the basis of population-scale phenomenological assumptions, one wants to derive them from first principles, that is, as the deterministic continuum limits of stochastic discrete models, \emph{i.e.} individual-based (IB) models, which track the dynamics of single cells~\cite{anderson2007single,van2015simulating}. This is to ensure that the terms comprised in the model equations provide a faithful mean-field representation of the underlying cellular dynamics. In fact, although being computationally intensive to simulate for large cell numbers and, to a wider extent, inaccessible to analytical techniques, IB models permit the representation of the finer details of cell-scale mechanisms and capture stochastic intercellular variability in the spatial and evolutionary trajectories of single cells. These aspects, which cannot be directly incorporated into phenomenological deterministic continuum models, become especially important in scenarios where cell numbers and densities are low (\emph{e.g.} in the early stages of embryonic development and tissue regeneration, during the formation of distant metastases upon cancer cell extravasation, and when tumour size is severely reduced after therapy), due to the stronger impact that single-cell processes and demographic stochasticity are expected to have on the dynamics of cell populations. For this reason, a range of asymptotic techniques, probabilistic methods and limiting procedures have been developed and used in previous studies to systematically derive PDE models for the growth of cell populations from their individual-based counterparts~\cite{lorenzi2022cancer}. {{For example, reaction-diffusion and nonlinear diffusion equations have been derived from their underlying random walks~\cite{inoue1991derivation,oelschlager1989derivation,penington2011building,chaplain2020bridging,lorenzi2020individual,baker2019free}, from systems of discrete equations of motion~\cite{oelschlager1990large,murray2009discrete,murray2012classifying}, from discrete lattice-based exclusion processes~\cite{dyson2012macroscopic,johnston2017co,johnston2012mean,johnston2015modelling} and from cellular automata~\cite{deroulers2009modeling,drasdo2005coarse,simpson2007simulating}.}}
However, these previous studies have mainly been focused on homogeneous populations in which all cells have the same phenotypic characteristics. Such homogeneity is rarely present in cellular systems, where significant intercellular phenotypic variability is commonly observed. In light of these considerations, we develop here an IB model for the growth of phenotypically heterogeneous cell populations. In this model, every cell is viewed as an individual agent whose phenotypic state is described by a structuring variable that captures intercellular variability in cell proliferation and migration rates. Cells undergo  directional movement in response to pressure differentials~\cite{ambrosi2002closure,byrne1997free,byrne2009individual,greenspan1976growth}, pressure-dependent proliferation~\cite{bru2003universal,byrne2003modelling,drasdo2012modeling,ranft2010fluidization}, and heritable phenotypic changes~\cite{brock2009non,chisholm2016cell,huang2013genetic} according to a set of rules that correspond to a discrete-time branching random walk on the physical space and the space of phenotypic states~\cite{chaplain2020bridging,chisholm2016evolutionary,hughes1995random}. We formally show that the deterministic continuum limit of this model is given by a non-local PDE for the cell population density function, which generalises earlier models~\cite{byrne2009individual,drasdo2012modeling,perthame2014hele} to the case of phenotypically heterogeneous cell populations. We then carry out numerical simulations of the IB model and compare the results obtained with the results of formal travelling-wave analysis and numerical simulations of the PDE model.

The paper is organised as follows. In Section~\ref{sec:model}, we introduce the IB model. In Section~\ref{sec:formal}, we present its PDE counterpart (a formal derivation is provided in Appendix A). In Section~\ref{sec:numerics}, in order to obtain results with broad structural stability under parameter changes, we first carry out formal travelling-wave analysis of the PDE model and then integrate the results obtained with numerical simulations of the IB model and numerical solutions of the PDE model. In Section~\ref{sec:conclusion}, we summarise the main findings of our study and outline directions for future research.

%%%%%%%%%%%%%%%%%%%%%%%%%%%%%%%%%%%%%%%%%
%% Model %% 
%%%%%%%%%%%%%%%%%%%%%%%%%%%%%%%%%%%%%%%%%
\section{The individual-based model}
\label{sec:model}
We model the dynamics of a phenotypically heterogeneous growing cell population. Cells within the population have the potential to undergo: 
\begin{itemize}
\item[(i)]  directional movement in response to pressure differentials -- \emph{i.e.} cells move down pressure gradients towards regions where they feel less compressed~\cite{ambrosi2002closure,byrne2003modelling,byrne2009individual};
\item[(ii)] spontaneous, heritable phenotypic changes, which lead cells to randomly transition from one phenotypic state into another~\cite{brock2009non,chisholm2016cell,huang2013genetic};
\item[(iii)] pressure-dependent proliferation -- \emph{i.e.} cells stop dividing, and can thus only die or remain quiescent, when the pressure that they experience overcomes a critical threshold, which is known as homeostatic pressure~\cite{basan2009homeostatic,byrne2009individual,drasdo2012modeling}.
\end{itemize}
Focussing on a one-dimensional spatial domain scenario, the position of every cell at time $t\in \mathbb{R}^{+}$ is described by the variable $x \in \mathbb{R}$. Moreover, the phenotypic state of each cell is characterised by a structuring variable $y \in [0,Y] \subset \mathbb{R}^{+}$, with $Y>0$, which takes into account intercellular variability in cell proliferation and migration rates.
{{Here, the variable $y$ could represent the level of expression of a gene that regulates both cell division and cell migration, such as those involved in the epithelial-to-mesenchymal transition promoting tumour invasion~\cite{novikov2021mutational,alfonso2017biology,giese2003cost}. More specifically, the overexpression of some cancer-promoting genes has been shown to inhibit cell proliferation and promote cell migration in cancer cells, for example, FBXL10 expession in ovarian cancer cell lines~\cite{yan2018mir} and EphB2 expression in glioblastomas~\cite{wang2012ephb2}. Similarily, the downregulation of miR-451 observed in glioblastomas has been shown to reduce the proliferation rate and increase the migration potential of the cells~\cite{godlewski2010microrna}}}.

Therefore in the model, without loss of generality, we consider the case where larger values of $y$ correlate with a higher cell migration rate but a lower proliferation rate due to proliferation-migration tradeoffs~\cite{aktipis2013life,alfonso2017biology,gallaher2019impact,gerlee2009evolution,gerlee2012impact,giese2003cost,giese1996dichotomy,hatzikirou2012go,orlando2013tumor,pham2012density}.  

We discretise the time, space and phenotype variables via 
$$
t_k=k \tau \in \mathbb{R}^{+}, \;\; x_i=i \chi \in \mathbb{R} \; \text{ and } \; y_j=j \eta \in [0,Y] \; \text{ with } \;  k, j \in \mathbb{N}_0, \; i \in \mathbb{Z}, \; \tau, \chi, \eta \in \mathbb{R}^{+}_*.
$$
Here $\tau$, $\chi$ and $\eta$ are the time-, space- and phenotype-step, respectively. We represent every single cell as an agent that occupies a position on the lattice $\{x_i\}_{i \in \mathbb{Z}} \times \{y_j\}_{j \in \mathbb{N}_0}$, and we introduce the dependent variable $N_{i,j}^k\in\mathbb{N}_0$ to model the number of cells in the phenotypic state $y_j$ at position $x_i$ at time $t_k$. The cell population density and the corresponding cell density are defined, respectively, as follows
\begin{equation}
 n_{i,j}^{k} \equiv n(t_k,x_i,y_j) := \frac{N_{i,j}^{k}}{\chi\eta} \quad \text{and} \quad \rho_{i}^{k} \equiv \rho(t_k,x_i) := \eta \sum_{j} n_{i,j}^{k}.\label{eq:density_define}
\end{equation}
We further define the pressure experienced by the cells (\emph{i.e.} the cell pressure) as a function of the cell density through the following barotropic relation
\begin{equation}
p_i^k  \equiv p(t_k,x_i) = \Pi(\rho_{i}^{k}),\label{eq:pressure_define}
\end{equation}
where the function $\Pi$ satisfies the following assumptions~\cite{byrne2009individual,perthame2014hele,tang2014composite} 
\begin{equation}
\Pi(0)=0,\quad \frac{\mathrm{d}}{\mathrm{d}\rho} \Pi(\rho) \geq 0 \; \text{ for }\; \rho \in \mathbb{R}^+_*.
\label{pressure_cond}
\end{equation}

\begin{figure}[tbhp]
\centering\includegraphics[width=\textwidth]{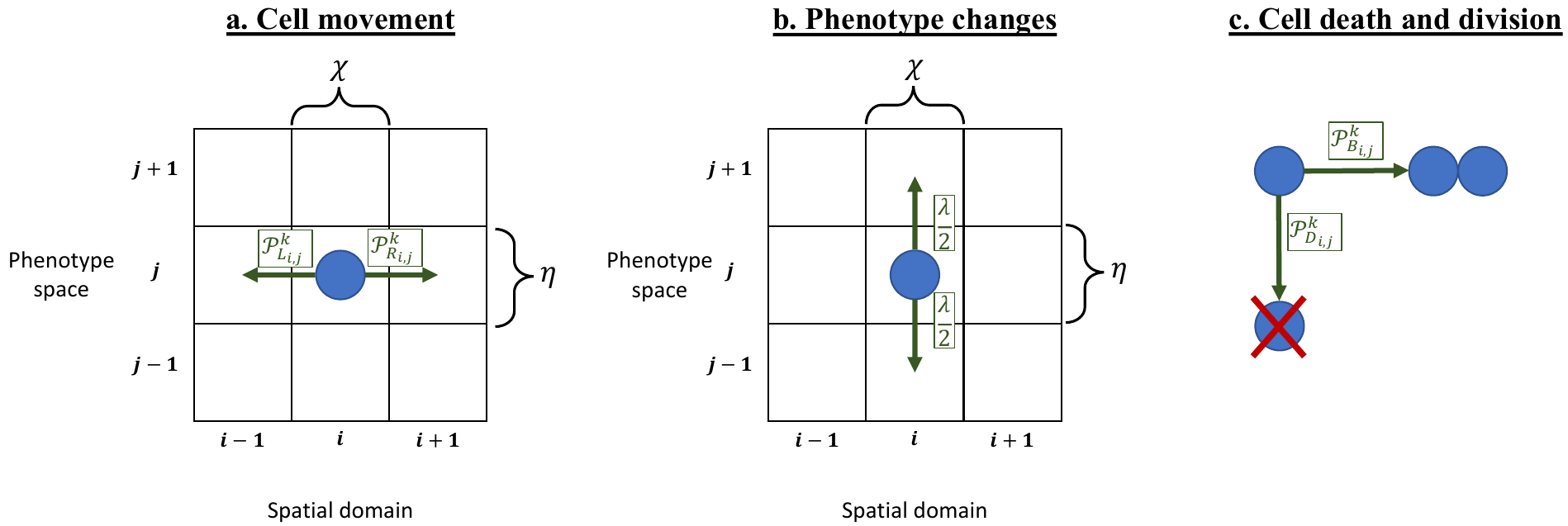}
\caption{Schematics summarising the rules that govern the spatial evolutionary dynamics of single cells in the IB model. Between time-steps $k$ and $k+1$, each cell in phenotypic state $y_j \in (0,Y)$ at position $x_i \in \mathbb{R}$ may: \textbf{a.} move to either of the positions $x_{i-1}$ and $x_{i+1}$ with probabilities $\euscr{P}^k_{L_{i,j}}$ and $\euscr{P}^k_{R_{i,j}}$ defined via~\eqref{Prob_mov_left}~and~\eqref{Prob_mov_right}; \textbf{b.} undergo a phenotypic change and thus enter into either of the phenotypic states $y_{j-1}$ and $y_{j+1}$ with probabilities $\lambda/2$; \textbf{c.} die or divide with probabilities $\euscr{P}^k_{D_{i,j}}$ and $\euscr{P}^k_{B_{i,j}}$ defined via~\eqref{Prob_death}~and~\eqref{Prob_birth}.}\label{fig:diagram}
\end{figure}

As summarised by the schematics in Figure~\ref{fig:diagram}, between time-steps $k$ and $k+1$, each cell in phenotypic state $y_j \in (0,Y)$ at position $x_i \in \mathbb{R}$ can first move, next undergo phenotypic changes and then die or divide according to the rules described in the following subsections.

\subsection{Mathematical modelling of cell death and division}
To incorporate the effects of cell proliferation, we assume that a dividing cell is { instantly} replaced by two identical cells that inherit the phenotypic state of the parent cell (\emph{i.e.} the progenies are placed on the same lattice site as their parent), while a dying cell is { instantly} removed from the population. We model pressure-dependent proliferation by letting the cells divide, die or remain quiescent with probabilities that depend on their phenotypic states and the pressure that they experience. In particular, to define the probabilities of cell division and death, we introduce the function $R(y_j, p_i^k)$, which describes the net growth rate of the cell population density at position $x_i$ at time $t_k$, and assume that between time-steps $k$ and $k+1$ a cell in phenotypic state $y_j$ at position $x_i$ may die with probability
\begin{equation}
\euscr{P}_{D_{i,j}}^{k} := \tau R(y_j,p^{k}_{i})_- \; \text{ where } \; R(y_j,p^{k}_{i})_- =  -\min\left(0,R(y_j,p^{k}_{i})\right),
\label{Prob_death}
\end{equation}
divide with probability    
\begin{equation}
\euscr{P}_{B_{i,j}}^{k} := \tau R(y_j,p^{k}_{i})_+ \; \text{ where } \; R(y_j,p^{k}_{i})_+ = \max\left(0,R(y_j,p^{k}_{i})\right)
\label{Prob_birth}
\end{equation}
or remain quiescent (\emph{i.e.} do not divide nor die) with probability
\begin{equation}
\euscr{P}_{Q_{i,j}}^{k} := 1-\euscr{P}_{B_{i,j}}^{k}-\euscr{P}_{D_{i,j}}^{k}.
\label{Prob_quiescent}
\end{equation}
Note that  we are implicitly assuming the time-step $\tau$ to be sufficiently small that $0 < \euscr{P}_{B_{i,j}}^{k} + \euscr{P}_{D_{i,j}}^{k} < 1$ for all values of $i$, $j$ and $k$. 

In order to capture the fact that, as mentioned earlier, larger values of $y_j$ correlate with a lower cell proliferation rate, along with the fact that cells will stop dividing if the pressure at their current position becomes larger than the homeostatic pressure, which we model by means of the parameter $p_M>0$, we make the following assumptions 
\begin{equation}
 \label{ass:R}
R(Y,0) = 0, \; R(0,p_M) = 0, \; \partial_p R(y,p) < 0 \; \text{ and } \; \partial_y R(y,p) < 0 \; \text{ for } (y,p) \in (0,Y) \times \mathbb{R}^+.
\end{equation}
In particular, we will focus on the case where
\begin{equation}
 \label{def:R}
R(y,p) := r(y) - \dfrac{p}{p_M} \; \text{ with } \; r(Y)=0, \;\; r(0)=1, \;\; \frac{\mathrm{d}}{\mathrm{d} y}r(y)<0 \; \text{for} \; y \in (0,Y).
\end{equation}

\begin{remark}
Under assumptions~\eqref{ass:R}, definitions~\eqref{Prob_birth}-\eqref{Prob_quiescent} ensure that if $p^{k}_{i} \geq p_M$ then every cell at position $x_i$ can only die or remain quiescent between time-steps $k$ and $k+1$. Hence, in the remainder of the paper we will let the following condition hold
\begin{equation}
\label{ass:p0IB}
\max_{i\in\mathbb{Z}} p_i^0 \leq p_M
\end{equation}
so that
\begin{equation}
p^{k}_{i}\le p_M \; \text{ for all } \; (k,i) \in \mathbb{N}_0 \times \mathbb{Z}. \label{bar_p}
\end{equation}
\end{remark}

\subsection{Mathematical modelling of phenotypic changes}
\label{phenotype_switch}
We  take into account heritable phenotypic changes by allowing cells to update their phenotypic states according to a random walk along the phenotypic dimension. More precisely, between time-steps $k$ and $k+1$, every cell either enters a new phenotypic state, with probability $\lambda \in [0,1]$, or remains in its current phenotypic state, with probability $1-\lambda$. Since, as mentioned earlier, we consider only spontaneous phenotypic changes that occur randomly due to non-genetic instability, we assume that a cell in phenotypic state $y_j$ that undergoes a phenotypic change enters into either of the phenotypic states $y_{j\pm1} = y_{j} \pm \eta$ with probabilities $\lambda/2$. No-flux boundary conditions are implemented by aborting any attempted phenotypic variation of a cell if it requires moving into a phenotypic state outside the interval $[0,Y]$.

\subsection{Mathematical modelling of cell movement}
We model directional cell movement in response to pressure differentials as a biased random walk along the spatial dimension, whereby the movement probabilities depend on the difference between the pressure at the position occupied by a cell and the pressure at the neighbouring positions. As mentioned earlier, we consider the case where larger values of $y_j$ correlate with a higher cell migration rate. Hence, we modulate the probabilities of cell movement by the function $\mu(y_j)$, which provides a measure of the mobility of cells in phenotypic state $y_j$ and thus satisfies the following assumptions
\begin{equation}
\mu(0)>0, \quad \frac{\mathrm{d}}{\mathrm{d}y}\mu(y)>0 \; \text{ for } \; y \in (0,Y].
\label{mu_cond}
\end{equation} 
Then we assume that between time-steps $k$ and $k+1$ a cell in phenotypic state $y_j$ at position $x_i$ may move to the position $x_{i-1} = x_i - \chi$ (\emph{i.e.} move left) with probability
\begin{equation}
\euscr{P}_{L_{i,j}}^{k}=\nu \mu(y_j)\frac{\left(p^{k}_{i}-p^{k}_{i-1}\right)_+}{2 p_M} \; \text{ where } \; \left(p^{k}_{i}-p^{k}_{i-1}\right)_+ =\max\left(0,p^{k}_{i}-p^{k}_{i-1}\right),
\label{Prob_mov_left}
\end{equation}
move to the position $x_{i+1} = x_i + \chi$ (\emph{i.e.} move right) with probability
\begin{equation}
\euscr{P}_{R_{i,j}}^{k}=\nu \mu(y_j)\frac{\left(p^{k}_{i}-p^{k}_{i+1}\right)_+}{2 p_M} \; \text{ where } \; \left(p^{k}_{i}-p^{k}_{i+1}\right)_+ =\max\left(0,p^{k}_{i}-p^{k}_{i+1}\right)
\label{Prob_mov_right}
\end{equation}
or remain stationary (\emph{i.e.} do not move left nor right) with probability
\begin{equation}
\euscr{P}_{S_{i,j}}^{k}=1-\euscr{P}_{L_{i,j}}^{k}-\euscr{P}_{R_{i,j}}^{k}.
\label{Prob_mov_stay}
\end{equation}
Here, the parameter $\nu>0$ is a scaling factor, which we implicitly assume to be sufficiently small that $0 < \nu \, \mu(y_j) < 1$ for all $y_j \in [0,Y]$. Under condition~\eqref{ass:p0IB}, this assumption on $\nu$ along with the a priori estimate~\eqref{bar_p} implies that definitions~\eqref{Prob_mov_left} and~\eqref{Prob_mov_right} are such that $0 < \euscr{P}_{L_{i,j}}^{k} + \euscr{P}_{R_{i,j}}^{k} < 1$ for all values of $i$, $j$ and $k$. 

\begin{remark}
Definitions~\eqref{Prob_mov_left} and~\eqref{Prob_mov_right} ensure that cells will move down pressure gradients so as to reach regions where they feel less compressed.
\end{remark}

\section{The corresponding continuum model}
\label{sec:formal}
Through a method analogous to those that we previously employed in~\cite{ardavseva2020comparative,bubba2020discrete,chaplain2020bridging,macfarlane2020hybrid,stace2020discrete}, letting the time-step $\tau\rightarrow 0$, the space-step $\chi\rightarrow 0$ and the phenotype-step $\eta\rightarrow0$ in such a way that
\begin{equation}
\frac{\nu\chi^2 }{2\tau}\rightarrow \alpha \in \mathbb{R}^+_* \quad \text{and} \quad \frac{\lambda \eta^2}{2\tau}\rightarrow \beta \in \mathbb{R}^+_*,
\label{derived_parameters}
\end{equation}
one can formally show (see Appendix A) that the deterministic continuum counterpart of the stochastic discrete model presented in Section~\ref{sec:model} is given by the following non-local PDE for the cell population density function $n(t,x,y)$ 
\begin{equation}
\begin{cases}
\displaystyle{ \, \partial_t n - \alpha \, \hat{\mu}(y) \, \partial_x \left(n \, \partial_x p \right) = R(y,p) \, n + \beta \, \partial^2_{yy} n, \quad (x,y) \in \mathbb{R} \times (0,Y)}
\\\\
\displaystyle{p=\Pi(\rho), \quad \rho:=\int_{0}^{Y} n(t,x,y) \ \mathrm{d} y,}
\end{cases}
\label{derived_PDE}
\end{equation}
where $\hat{\mu}(y) := \dfrac{\mu(y)}{p_M}$. The non-local PDE~\eqref{derived_PDE} is subject to zero Neumann (\emph{i.e.} no-flux) boundary conditions at $y=0$ and $y=Y$, as well as to an initial condition such that the continuum analogue of condition~\eqref{ass:p0IB} holds, that is, 
\begin{equation}
\label{ass:p0PDE}
\max_{x \in \mathbb{R}} p(0,x) \leq p_M.
\end{equation} 

The mathematical model defined by complementing~\eqref{derived_PDE} with assumptions~\eqref{pressure_cond}, \eqref{ass:R} and~\eqref{mu_cond} generalises earlier models of pressure-dependent cell population growth~\cite{byrne2009individual,drasdo2012modeling,perthame2014hele} to the case of phenotypically heterogeneous cell populations. 

%%%%%%%%%%%%%%%%%%%%%%%%%%%%%%%%%%%%%%%%%
%% Results %% 
%%%%%%%%%%%%%%%%%%%%%%%%%%%%%%%%%%%%%%%%%
\section{Main results}
\label{sec:numerics}
In this section, we first present the result of formal travelling-wave analysis of the PDE model (Subsection~\ref{Sec4}) and then integrate these results with numerical simulations of the IB model and numerical solutions of the PDE model (Subsection~\ref{Sec:numres}). 

\subsection{Formal travelling-wave analysis}
\label{Sec4}
We focus on a biological scenario in which cell movement occurs on a slower timescale compared to cell division and death, while spontaneous, heritable phenotypic changes occur on a
slower timescale compared to cell movement~\cite{huang2013genetic,smith2004measurement}. To this end, we introduce a small parameter $\varepsilon >0$ and let 
\begin{equation}
\label{eq:abeps}
\alpha := \varepsilon, \quad \beta := \varepsilon^2.
\end{equation}
Moreover, in order to explore the long-time behaviour of the cell population (\emph{i.e.} the behaviour of the population over many cell generations), we use the time scaling $t \to \frac{t}{\varepsilon}$ in~\eqref{derived_PDE}. Taken together, this gives the following non-local PDE for the cell population density function $n_{\varepsilon}(t,x,y) = n(\frac{t}{\varepsilon},x,y)$ 
\begin{equation}
\begin{cases}
\displaystyle{\varepsilon \, \partial_t n_{\varepsilon} - \varepsilon \, \hat{\mu}(y) \, \partial_x \left(n_{\varepsilon} \, \partial_x p_{\varepsilon} \right) = R(y,p_{\varepsilon}) \, n_{\varepsilon} + \varepsilon^2 \, \partial^2_{yy} n_{\varepsilon}, \quad (x,y) \in \mathbb{R} \times (0,Y)}
\\\\
\displaystyle{p_{\varepsilon}=\Pi(\rho_{\varepsilon}), \quad \rho_{\varepsilon}:=\int_{0}^{Y} n_{\varepsilon}(t,x,y) \ \mathrm{d} y.}
\end{cases}
\label{eq:PDEnen}
\end{equation}
Using a method analogous to those that we have previously employed in~\cite{lorenzi2021,lorenzi2022trade}, denoting by $\delta_{(\cdot)}(y)$ the Dirac delta centred at $y=(\cdot)$, one can formally show (see Appendix B) that, under assumptions~\eqref{pressure_cond}, \eqref{def:R} and~\eqref{mu_cond}, as $\varepsilon \to 0$, the non-local PDE~\eqref{eq:PDEnen} admits travelling-wave solutions of the form
\begin{equation}
\label{eq:concefn}
n_{\varepsilon}(z,y) \approx \rho(z) \, \delta_{\bar{y}(z)}(y), \quad z = x - c \, t, \quad c \in \mathbb{R}^+_*,  
\end{equation}
where $\bar{y}(z)$ is the unique maximum point of the solution to the following equation
$$
- \left(c + \hat{\mu}(y) p' \right) \partial_z u = \left(r(y) - \dfrac{p}{p_M} \right) + (\partial_y u)^2, \quad u \equiv u(z,y), \quad(z,y) \in \mathbb{R} \times (0,Y)
$$
subject to the constraint $\displaystyle{\max_{y \in [0,Y]} u(z,y) = u(z,\overline{y}(z)) = 0}$ for $z \in \supp(\rho)$, and the cell density $\rho(z)$ is such that the pressure $p(z) = \Pi(\rho(z))$ satisfies the following relation 
\begin{equation}
\label{eq:expptwpap}
p(z) = p_M \, r\left(\bar{y}(z)\right) \quad z \in \supp(p),
\end{equation}
provided that the wave speed $c$ satisfies the following necessary condition 
\begin{equation}
\label{eq:cminex}
c \geq \sup_{z \in {\rm Supp}(r(\bar{y}))} 2 \ \left|\dfrac{{\rm d} }{{\rm d} y} r(\bar{y}(z))\right| \sqrt{\dfrac{\mu(\bar{y}(z))}{\left|\partial^2_{yy} u(z,\bar{y}(z))\right|}}.
\end{equation}
Moreover, 
\begin{equation}
\label{eq:TWpsuppap}
\supp(p) = (-\infty, \ell) \; \text{ with } \; \ell \in \mathbb{R} \; \text{ such that } \; \bar{y}(\ell) = Y
\end{equation} 
and
\begin{equation}
\label{eq:TWmonbarypap}
\lim_{z \to - \infty} \bar{y}(z) =0, \quad \lim_{z \to - \infty} p(z) =p_M, \quad \bar{y}'(z) > 0 \; \text{ and } \; p'(z) < 0 \quad z \in (-\infty, \ell).
\end{equation}

From a biological point of view, $\bar{y}(z)$ represents the dominant phenotype of cells at a certain position along the invading wave $p(z)$. Since larger values of $y$ correlate with a lower proliferation rate and a higher migration rate, the fact that $\bar{y}(z)$ increases monotonically from $0$ to $Y$ while $p(z)$ decreases monotonically from $p_M$ to $0$ (cf. the results given by~\eqref{eq:TWpsuppap} and~\eqref{eq:TWmonbarypap}) provides a mathematical formalisation of the idea that spatial sorting causes cells with a more mobile/less proliferative phenotype to become concentrated towards the front of the invading wave, which is thus a sparsely populated region, whereas phenotypic selection leads cells with a less mobile/more proliferative phenotype to dominate at the rear, which is then a densely populated region. 

\subsection{Numerical simulations}
\label{Sec:numres}
\subsubsection{Set-up of numerical simulations}
In order to carry out numerical simulations, we consider the time interval $[0,T]$, with $T=8$, we restrict the physical domain to the closed interval $[0, X]$, with $X=25$, and choose $Y=1$. In order to facilitate the integration between numerical simulations and the results of formal travelling-wave analysis presented in Subsection~\ref{Sec4}, we solve numerically the rescaled PDE model~\eqref{eq:PDEnen}, with $\varepsilon = 0.01$, and we carry out numerical simulations of the scaled IB model obtained by introducing the time scaling $t_k \to \dfrac{t_k}{\varepsilon} = k \dfrac{\tau}{\varepsilon}$ and reformulating the governing rules of cell dynamics that are detailed in Section~\ref{sec:model} in terms of 
$$
p_{\varepsilon i}^k \equiv p_{\varepsilon}(t_k,x_i) = p\left(\frac{t_k}{\varepsilon},x_i\right) = \Pi(\rho_{\varepsilon i}^{k}),
$$
with
$$
\rho_{\varepsilon i}^{k} \equiv \rho_{\varepsilon}(t_k,x_i) = \rho\left(\frac{t_k}{\varepsilon},x_i\right) := \eta \sum_{j} n_{\varepsilon i,j}^{k} \quad \text{and} \quad n_{\varepsilon i,j}^{k} \equiv n_{\varepsilon}\left(t_k, x_i, y_j \right) = n\left(\frac{t_k}{\varepsilon}, x_i, y_j \right) := \frac{N_{\varepsilon i,j}^{k}}{\chi\eta}.
$$
{{Moreover, we choose $\tau =  5\times10^{-5}$, $\chi=0.01$ and $\eta=0.02$, and then set $\nu = \dfrac{2 \tau}{\chi^2} \varepsilon$ and $\lambda = \dfrac{2 \tau}{\eta^2} \varepsilon^2$ in order to ensure that conditions~\eqref{derived_parameters} and~\eqref{eq:abeps} are simultaneously satisfied.}}

We consider a biological scenario in which, initially, the cell population is localised along the $x=0$ boundary and most of the cells are in the phenotypic state $y=\bar{y}^0$ at every position. Specifically, we implement the following initial cell distribution for the IB model
\begin{equation}
\label{def:icIB}
N_{\varepsilon i,j}^0=\text{int}(F_{\varepsilon}(x_i,y_j)) \; \text{ with } \; F_{\varepsilon}(x,y)= A_0 \ C\ e^{-x^2} \ e^{-\frac{\left(y-\bar{y}^0\right)^2}{\varepsilon}},
\end{equation}
where $\text{int}(\cdot)$ is the integer part of $(\cdot)$ and $C$ is a normalisation constant such that 
$$
C\int_0^Y e^{-\frac{\left(y-\bar{y}^0\right)^2}{\varepsilon}} \ \mathrm{d}y=1.
$$
Unless otherwise specified, we choose $A_0=10$ and $\bar{y}^0=0.2$, {{that is, the initially dominant phenotype of the cell population is $y=0.2$}}. The initial cell density and pressure are then calculated via~\eqref{eq:density_define} and~\eqref{eq:pressure_define}. The initial cell population density function $n_\varepsilon(0,x,y) = n^0_\varepsilon(x,y)$, is defined as a suitable continuum analogue of the cell population density $n_{\varepsilon i,j}^{0} := \frac{N_{\varepsilon i,j}^{0}}{\chi\eta}$, with $N_{\varepsilon i,j}^{0}$ given by~\eqref{def:icIB}. Specifically, we set
$$
n^0_\varepsilon(x,y) := \left(\frac{F_{\varepsilon}(x,y)}{\chi \eta} - \dfrac{1.5}{\chi \eta}\right)_+,
$$
where $F_{\varepsilon}(x,y)$ is defined via~\eqref{def:icIB} and $(\cdot)_{+}$ is the positive part of $(\cdot)$. 

We define $R(y,p_{\varepsilon})$ via~\eqref{def:R} and, having chosen $Y=1$, we further define 
$$
r(y) :=1-y^2 \quad \text{and} \quad \mu(y) := 0.01 + y^2
$$
so as to ensure that assumptions~\eqref{ass:R}, \eqref{def:R} and~\eqref{mu_cond} are satisfied. Moreover, we investigate the following three definitions of the barotropic relation for the cell pressure, all satisfying assumptions~\eqref{pressure_cond}:
\begin{equation}
p_{\varepsilon} = \Pi(\rho_{\varepsilon}) :=
\begin{cases}
\quad \rho_{\varepsilon} & \text{(Case 1)}\\
&\\
\quad K_{\gamma} \left(\rho_{\varepsilon}\right)^\gamma \quad\quad\; \text{with} \quad K_{\gamma}>0, \ \gamma>1 & \text{(Case 2)}\\
&\\
\quad \kappa \left(\rho_{\varepsilon}-\rho^*\right)_+ \quad \text{with} \quad \kappa,\ \rho^*> 0 & \text{(Case 3)}.
\label{Pressure_all}
\end{cases}
\end{equation}
The definition given by Case 1 corresponds to the simplified scenario in which the cell pressure is a linear function of the cell density. In the definition given by Case 2, which was proposed in~\cite{perthame2014hele}, the parameter $K_{\gamma}$ is a scaling factor and the parameter $\gamma$ provides a measure of the stiffness of the barotropic relation (\emph{i.e.} the limit $\gamma\rightarrow \infty$ corresponds to the scenario in which cells behave like an incompressible fluid). In the definition given by Case 3, which is such that the cell pressure is zero for $\rho \leq  \rho^*$ and is a monotonically increasing function of the cell density for $\rho > \rho^*$, the parameter $\kappa$ is a scaling factor and $\rho^*$ is the density below which the force that the cells exert upon one another is negligible~\cite{drasdo2012modeling,tang2014composite}. Unless otherwise specified: when the cell pressure is defined via Case 1 we choose $p_M= 4.95\times10^4$; when the cell pressure is defined via Case 2 we choose $p_M = 3.675\times10^9$, $\gamma=2$ and $K_{\gamma} = \frac{3}{2}$; when the cell pressure is defined via Case 3 we choose $p_M = 4.94\times10^5$, $\kappa=10$ and $\rho^*= 10^2$. 

\begin{remark}
The initial conditions and the values of $p_M$ that are considered here are such that conditions~\eqref{ass:p0IB} and~\eqref{ass:p0PDE} are satisfied.
\end{remark}

\subsubsection{Computational implementation of the IB model}
All simulations are performed in {\sc Matlab}. At each time-step, each cell undergoes a three-phase process: (i) cell movement, according to the probabilities defined via~\eqref{Prob_mov_left}-\eqref{Prob_mov_stay}; (ii) phenotypic changes, with probabilities $\lambda/2$; (iii) division and death, according to the probabilities defined via~\eqref{Prob_birth}-\eqref{Prob_quiescent}. For each cell, during each phase, a random number is drawn from the standard uniform distribution on the interval $(0,1)$ using the built-in {\sc Matlab} function {\sc rand}. It is then evaluated whether this number is lower than the probability of the event occurring and if so the event occurs. Since $x_i \in [0,X]$,  the attempted movement of a cell is aborted if it requires moving out of the spatial domain. 

\subsubsection{Methods used to solve numerically the non-local PDE~\eqref{eq:PDEnen}}
Full details of the methods used to solve numerically the non-local PDE~\eqref{eq:PDEnen} posed on $(0,T] \times (0,X) \times (0,Y)$ and subject to suitable initial and boundary conditions are given in Appendix C.

\subsubsection{Results of numerical simulations} 
\begin{figure}[tbhp]
\centering
\bf{Case 1}\\
\vspace{1em}

\includegraphics[height=0.5\textwidth]{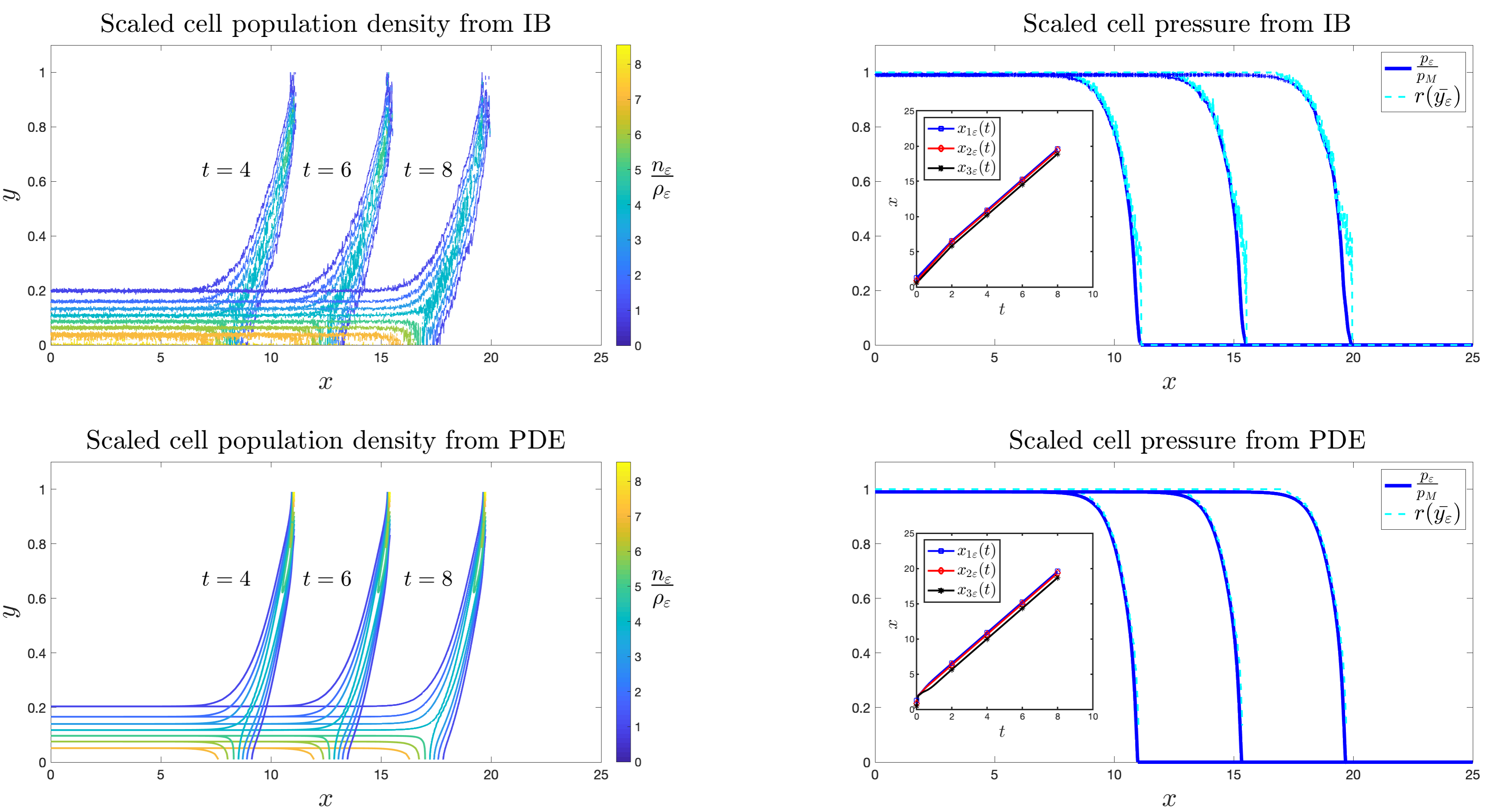}
\caption{Numerical simulation results of the IB model (top row) and the PDE model~\eqref{eq:PDEnen} (bottom row) in the case where the cell pressure is defined through the barotropic relation given by Case 1 in \eqref{Pressure_all}. Plots display the scaled cell population density $n_\varepsilon/\rho_\varepsilon$ (left panels) and the scaled cell pressure $p_\varepsilon/ p_M$ (right panels, solid blue lines) at three successive time instants (\emph{i.e.} $t = 4$, $t = 6$ and $t = 8$) for both modelling approaches. The dashed cyan lines in the right panels highlight the corresponding values of $r(\bar{y}_\varepsilon)$, with $\bar{y}_\varepsilon$ defined via~\eqref{def:yeps}. The insets of the right panels display the plots of $x_{\varepsilon 1}(t)$ (blue squares), $x_{\varepsilon 2}(t)$ (red diamonds) and $x_{\varepsilon 3}(t)$ (black stars) defined via~\eqref{def:xeps}. The results from the IB model were obtained by averaging over 10 simulations.}
\label{fig:test_P1}
\end{figure}

\begin{figure}[tbhp]
\centering
\bf{Case 2}\\
\vspace{1em}

\includegraphics[height=0.5\textwidth]{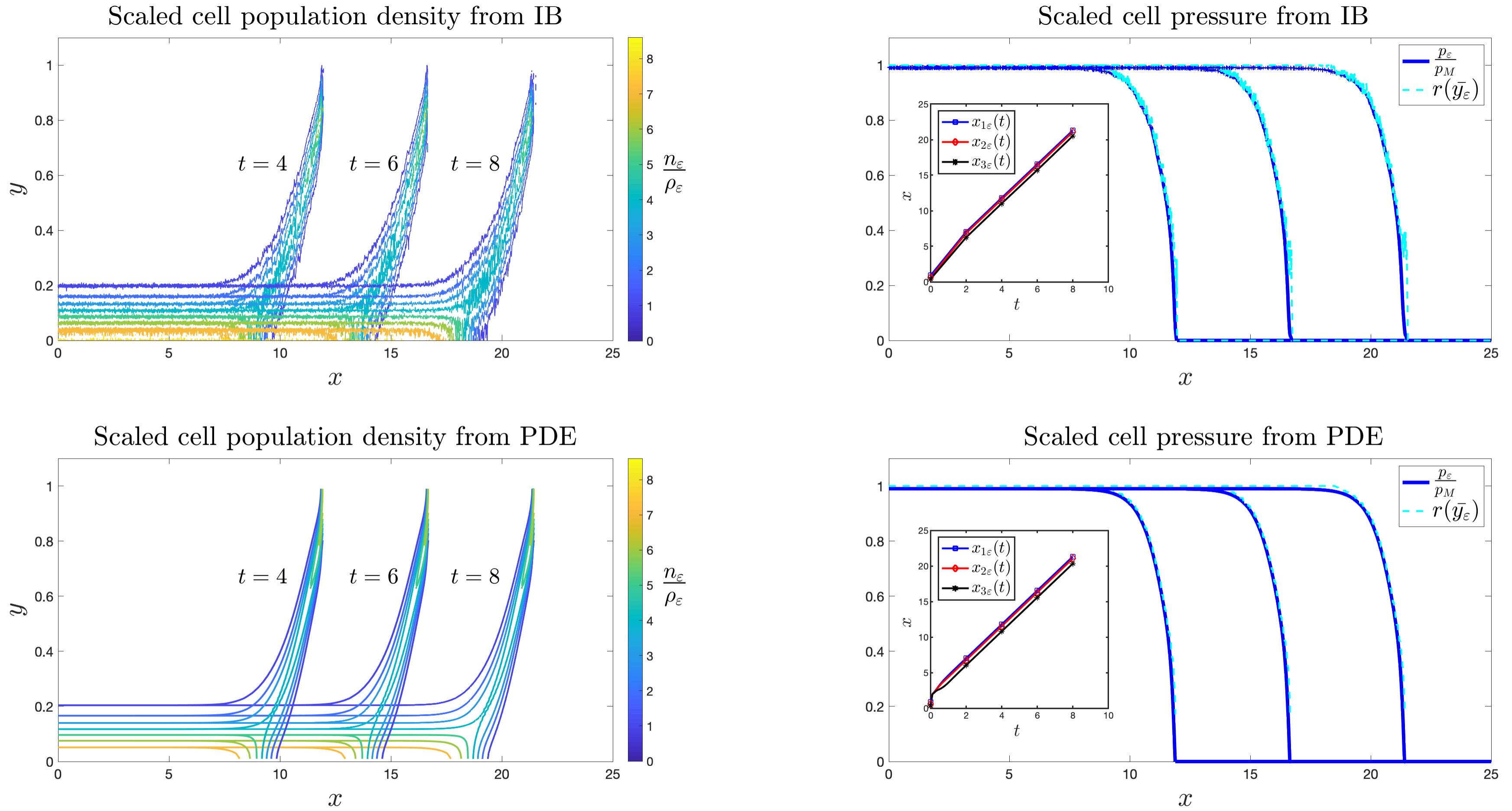}
\caption{Numerical simulation results of the IB model (top row) and the PDE model~\eqref{eq:PDEnen} (bottom row) in the case where the cell pressure is defined through the barotropic relation given by Case 2 in \eqref{Pressure_all}. Plots display the scaled cell population density $n_\varepsilon/\rho_\varepsilon$ (left panels) and the scaled cell pressure $p_\varepsilon/ p_M$ (right panels, solid blue lines) at three successive time instants (\emph{i.e.} $t = 4$, $t = 6$ and $t = 8$) for both modelling approaches. The dashed cyan lines in the right panels highlight the corresponding values of $r(\bar{y}_\varepsilon)$, with $\bar{y}_\varepsilon$ defined via~\eqref{def:yeps}. The insets of the right-hand panels display the plots of $x_{\varepsilon 1}(t)$ (blue squares), $x_{\varepsilon 2}(t)$ (red diamonds) and $x_{\varepsilon 3}(t)$ (black stars) defined via~\eqref{def:xeps}. The results from the IB model were obtained by averaging over 10 simulations.}
\label{fig:test_P2}
\end{figure}

\begin{figure}[tbhp]
\centering
\bf{Case 3}\\
\vspace{1em}

\includegraphics[height=0.5\textwidth]{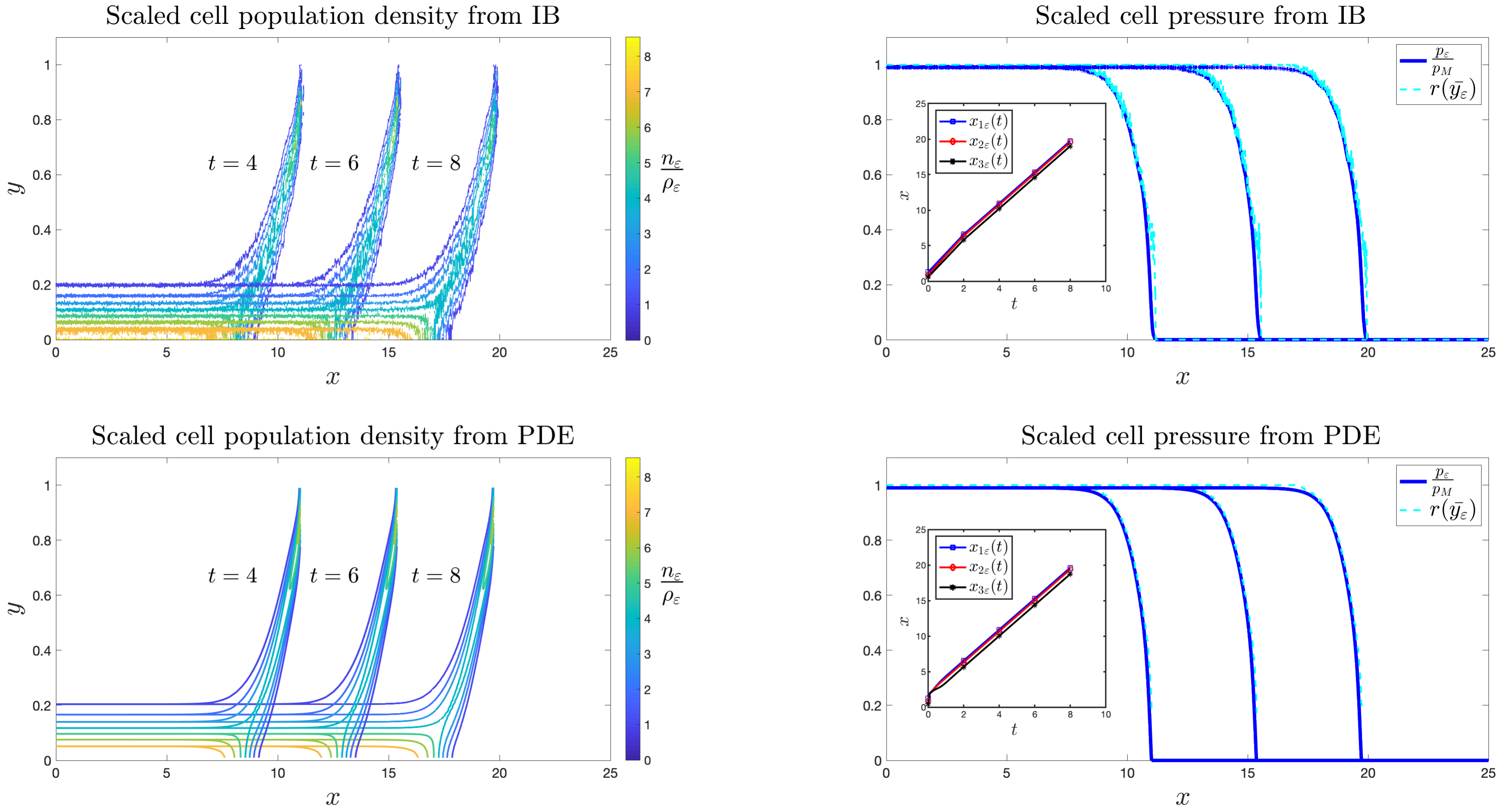}
\caption{Numerical simulation results of the IB model (top row) and the PDE model~\eqref{eq:PDEnen} (bottom row) in the case where the cell pressure is defined through the barotropic relation given by Case 3 in \eqref{Pressure_all}. Plots display the scaled cell population density $n_\varepsilon/\rho_\varepsilon$ (left panels) and the scaled cell pressure $p_\varepsilon/ p_M$ (right panels, solid blue lines) at three successive time instants (\emph{i.e.} $t = 4$, $t = 6$ and $t = 8$) for both modelling approaches. The dashed cyan lines in the right panels highlight the corresponding values of $r(\bar{y}_\varepsilon)$, with $\bar{y}_\varepsilon$ defined via~\eqref{def:yeps}. The insets of the right-hand panels display the plots of $x_{\varepsilon 1}(t)$ (blue squares), $x_{\varepsilon 2}(t)$ (red diamonds) and $x_{\varepsilon 3}(t)$ (black stars) defined via~\eqref{def:xeps}. The results from the IB model were obtained by averaging over 10 simulations.}
\label{fig:test_P3}
\end{figure}

\paragraph{Formation of complex spatial patterns of population growth.} The plots in the top lines of Figures~\ref{fig:test_P1}-\ref{fig:test_P3} summarise the results of numerical simulations of the IB model for the barotropic relations given by Cases 1-3 in~\eqref{Pressure_all}. We plot in the left panels the scaled cell population density, $n_{\varepsilon}/\rho_{\varepsilon}$, and in the right panels (solid blue lines) the scaled cell pressure, $p_{\varepsilon}/p_M$, at progressive times. 

The results of numerical simulations under all three cases display very similar dynamics, both with respect to the cell population density and the cell pressure, where we observe evolution to travelling-wave profiles of almost identical shapes and speeds (see also Remark~\ref{rem:robu}). The incorporation of proliferation-migration tradeoffs lead cells to be non-uniformly distributed across both physical and phenotype space. More precisely, we observe a relatively small subpopulation of highly-mobile but minimally-proliferative cells (\emph{i.e.} cells in phenotypic states $y\approx Y$) that becomes concentrated towards the front of the invading wave, while rapidly-proliferating but minimally-mobile cells (\emph{i.e.} cells in phenotypic states $y\approx 0$) make up the bulk of the population in the rear. This is due to a dynamic interplay between spatial sorting and phenotypic selection. In fact, a more efficient response to pressure differentials by more-mobile cells leads to their positioning at the front of the wave, where the pressure is lower, before being overcome by the more-proliferative cells encroaching from the rear of the wave, which are ultimately selected due to their higher proliferative potential.

\paragraph{Quantitative agreement between the IB model and its PDE counterpart.} The plots in the bottom lines of Figures~\ref{fig:test_P1}-\ref{fig:test_P3} summarise the corresponding numerical solutions of the PDE model~\eqref{eq:PDEnen}. Comparing these plots with those in the top lines, we can see that there is an excellent quantitative agreement between the results of numerical simulations of the IB model and the numerical solutions of its PDE counterpart, both with respect to the cell population density and the cell pressure, for each of the barotropic relations given by Cases 1-3 in~\eqref{Pressure_all}. 

All these plots indicate that, in agreement with the results presented in Subsection~\ref{Sec4} (cf. the results given by~\eqref{eq:concefn}), when $\varepsilon$ is sufficiently small, the scaled cell population density $n_\varepsilon/\rho_\varepsilon$ is concentrated as a sharp Gaussian with maximum at a single point $\bar{y}_{\varepsilon}(t,x)$ for all $x \in \supp(p_{\varepsilon})$. The maximum point $\bar{y}_{\varepsilon}(t,x)$ corresponds to the dominant phenotype within the cell population at position $x$ and time $t$. Again in agreement with the results presented in Subsection~\ref{Sec4} (cf. the results given by~\eqref{eq:TWpsuppap} and~\eqref{eq:TWmonbarypap}), the cell pressure $p_{\varepsilon}$ behaves like a one-sided compactly supported and monotonically decreasing travelling front that connects $p_M$ to $0$, while the dominant phenotype $\bar{y}_{\varepsilon}$ increases monotonically from $y=0$ to $y=Y$ across the support of the invading wave. Moreover, we find an excellent quantitative agreement between $p_{\varepsilon}(t,x)/p_M$ and $r(\bar{y}_{\varepsilon}(t,x))$ (cf. the solid blue and dashed cyan lines in the right panels of Figures~\ref{fig:test_P1}-\ref{fig:test_P3}). This indicates that, when $\varepsilon$ is sufficiently small, relation~\eqref{eq:expptwpap} is satisfied as well. 

In order to measure the speed of these travelling waves, we track the dynamics of the points $x_{\varepsilon 1}(t)$, $x_{\varepsilon 2}(t)$ and $x_{\varepsilon 3}(t)$ such that
\begin{equation}
\label{def:xeps}
p_{\varepsilon}(t, x_{\varepsilon 1}(t)) = 0.2 p_M, \quad p_{\varepsilon}(t, x_{\varepsilon 2}(t)) = 0.5 p_M, \quad p_{\varepsilon}(t, x_{\varepsilon 3}(t)) = 0.8 p_M.
\end{equation}
Notably, we observe the evolution of $x_{\varepsilon 1}(t)$, $x_{\varepsilon 2}(t)$ and $x_{\varepsilon 3}(t)$ towards straight lines of approximatively the same slope $\approx 2.5$ (see the insets of the right panels in Figures~\ref{fig:test_P1}-\ref{fig:test_P3}). Moreover, an equivalent tracking of $\tilde{x}_{\varepsilon 1}(t)$, $\tilde{x}_{\varepsilon 2}(t)$ and $\tilde{x}_{\varepsilon 3}(t)$ such that $\bar{y}_{\varepsilon}(t, \tilde{x}_{\varepsilon 1}(t)) = 0.2$, $\bar{y}_{\varepsilon}(t, \tilde{x}_{\varepsilon 2}(t)) = 0.5$ and $\bar{y}_{\varepsilon}(t, \tilde{x}_{\varepsilon 3}(t)) = 0.8$, with 
\begin{equation}
\label{def:yeps}
\bar{y}_\varepsilon(t,x) := \argmax_{y\in[0,Y]} n_\varepsilon(t,x,y),
\end{equation}
yields quasi-identical results (results not shown). This supports the idea that $p_{\varepsilon}$ behaves like a travelling front of speed $c \approx 2.5$. Such a value of the speed is coherent with the condition on the minimal wave speed given by~\eqref{eq:cminex}. In fact, inserting into~\eqref{eq:cminex} the numerical values of $\bar{y}_{\varepsilon}(8,x)$ in place of $\bar{y}(z)$ and the numerical values of $\partial^2_{yy} u_{\varepsilon}(8,x,\bar{y}_{\varepsilon}(8,x))$ with $u_{\varepsilon} = \varepsilon \log(n_{\varepsilon})$ in place of $\partial^2_{yy} u(z,\bar{y}(z))$ gives $c \gtrapprox 2.5$.

\begin{remark}
\label{rem:robu}
The robustness of the results of numerical simulations of the IB model presented so far is supported by the fact that there is an excellent quantitative agreement between them and the results of numerical simulations and formal travelling-wave analysis of the corresponding PDE model. In fact, in the light of this agreement, independently of the specific definitions of the model functions $\Pi$, $R$ and $\mu$, provided that assumptions~\eqref{pressure_cond}, \eqref{ass:R} and~\eqref{mu_cond} are satisfied, and sufficiently large cell numbers are considered, in the asymptotic regime $\varepsilon \to 0$, one can expect the rules governing the spatial evolutionary dynamics of single cells considered here to  bring about patterns of population growth that will ultimately be qualitatively similar to those of Figures~\ref{fig:test_P1}-\ref{fig:test_P3}.
\end{remark}

\paragraph{Possible discrepancies between the IB model and its PDE counterpart.} In the cases discussed so far, we have observed excellent quantitative agreement between {{averaged}} results of numerical simulations of the IB model and numerical solutions of the corresponding PDE model. However, we hypothesise that possible differences between the two models can emerge when cell dynamics are strongly impacted by demographic stochasticity, which cannot be captured by the PDE model. 

In general, we expect demographic stochasticity to have a stronger impact on cell dynamics in the presence of smaller values of the homeostatic pressure $p_M$, since smaller values of $p_M$ correlate with smaller cell numbers. Moreover, in the case where cells are initially distributed across both physical and phenotype space according to~\eqref{def:icIB}, we expect demographic stochasticity to escalate during the early stages of cell dynamics if sufficiently small values of the parameter $A_0$ and sufficiently large values of  the parameter $\bar{y}^0$ (\emph{i.e.} values of $\bar{y}^0$ sufficiently far from $0$ and sufficiently close to $Y$) are considered. In fact, smaller values of $A_0$ correlate with lower initial cell numbers. Moreover, since cells in phenotypic states $y\approx 0$ will ultimately be selected in the rear of the invading wave (cf. the plots in the left panels of Figures~\ref{fig:test_P1}-\ref{fig:test_P3}), bottleneck effects leading to a temporary drastic reduction in the size of the cell population may occur if $\bar{y}^0$ is sufficiently far from $0$. 

Hence, to test the aforementioned hypothesis, we carry out numerical simulations of the two models for decreasing values of $A_0$ and increasing values of $\bar{y}^0$ in the initial cell distribution~\eqref{def:icIB}. Furthermore, we define the cell pressure through the barotropic relation given by Case 1 in \eqref{Pressure_all} and set $\displaystyle{p_M := \max_{x \in [0,X]} \Pi\left(\rho_{\varepsilon}(0,x)\right)}$, so that smaller values of $A_0$ correspond to smaller value of $p_M$ as well.

\begin{figure}[tbhp]
\centering
\includegraphics[height=0.5\textwidth]{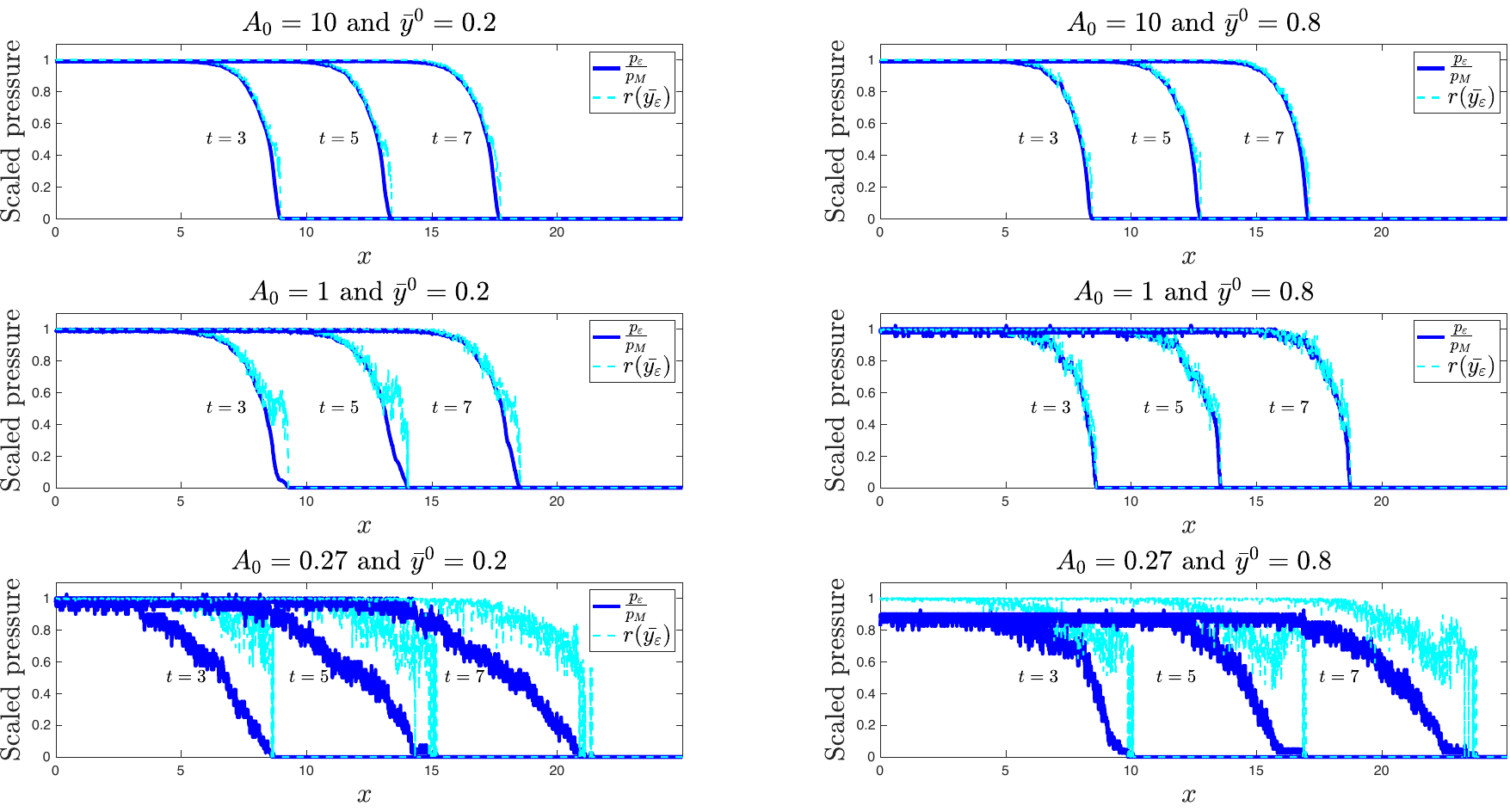}
\caption{Numerical simulation results of the IB model for different values of the parameters $A_0$ and $\bar{y}^0$ in the initial cell distribution~\eqref{def:icIB} -- \emph{i.e.} $\bar{y}^0=0.2$ (left column) or $\bar{y}^0=0.8$ (right column) and $A_0=10$ (top row) or $A_0=1$ (central row) or $A_0=0.27$ (bottom row). The solid blue lines highlight the values of the scaled cell pressure $p_\varepsilon/ p_M$ at three successive time instants (\emph{i.e.} $t = 3$, $t = 5$ and $t = 7$). The dashed cyan lines highlight the corresponding values of $r(\bar{y}_\varepsilon)$, with $\bar{y}_\varepsilon$ defined via~\eqref{def:yeps}. These results were obtained by averaging over 10 simulations.}
\label{fig:test_P4a}
\end{figure}

\begin{figure}[tbhp]
\centering
\includegraphics[height=0.5\textwidth]{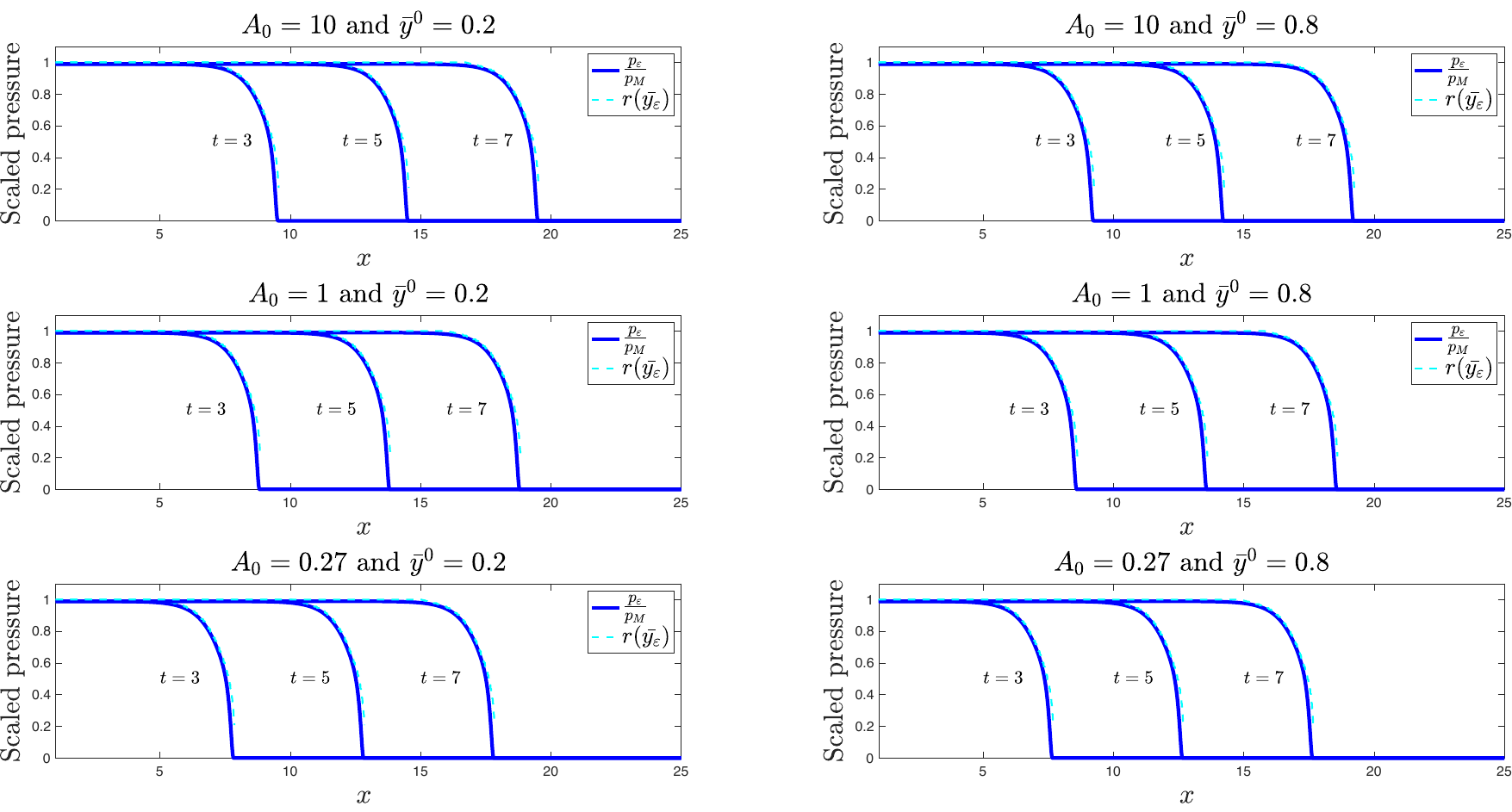}
\caption{Numerical simulation results of the PDE model~\eqref{eq:PDEnen} for different values of the parameters $A_0$ and $\bar{y}^0$ in the initial cell distribution~\eqref{def:icIB} -- \emph{i.e.} $\bar{y}^0=0.2$ (left column) or $\bar{y}^0=0.8$ (right column) and $A_0=10$ (top row) or $A_0=1$ (central row) or $A_0=0.27$ (bottom row). The solid blue lines highlight the values of the scaled cell pressure $p_\varepsilon/ p_M$ at three successive time instants (\emph{i.e.} $t = 3$, $t = 5$ and $t = 7$). The dashed cyan lines highlight the corresponding values of $r(\bar{y}_\varepsilon)$, with $\bar{y}_\varepsilon$ defined via~\eqref{def:yeps}.}
\label{fig:test_P4b}
\end{figure}

{{The results obtained for the IB model are summarised by the plots in Figure~\ref{fig:test_P4a}, which display typical dynamics of the scaled cell pressure $p_{\varepsilon}/p_M$, while the corresponding results for the PDE model~\eqref{eq:PDEnen} are summarised by the plots in Figure~\ref{fig:test_P4b}. These results corroborate our hypothesis by demonstrating that the quantitative agreement between the IB model and its PDE counterpart deteriorates when smaller values of $A_0$ and larger values of $\bar{y}^0$ are considered (cf. the plots in the bottom-line panels of Figures~\ref{fig:test_P4a} and~\ref{fig:test_P4b}).}} 

\begin{figure}[tbhp]
\centering
\includegraphics[height=0.5\textwidth]{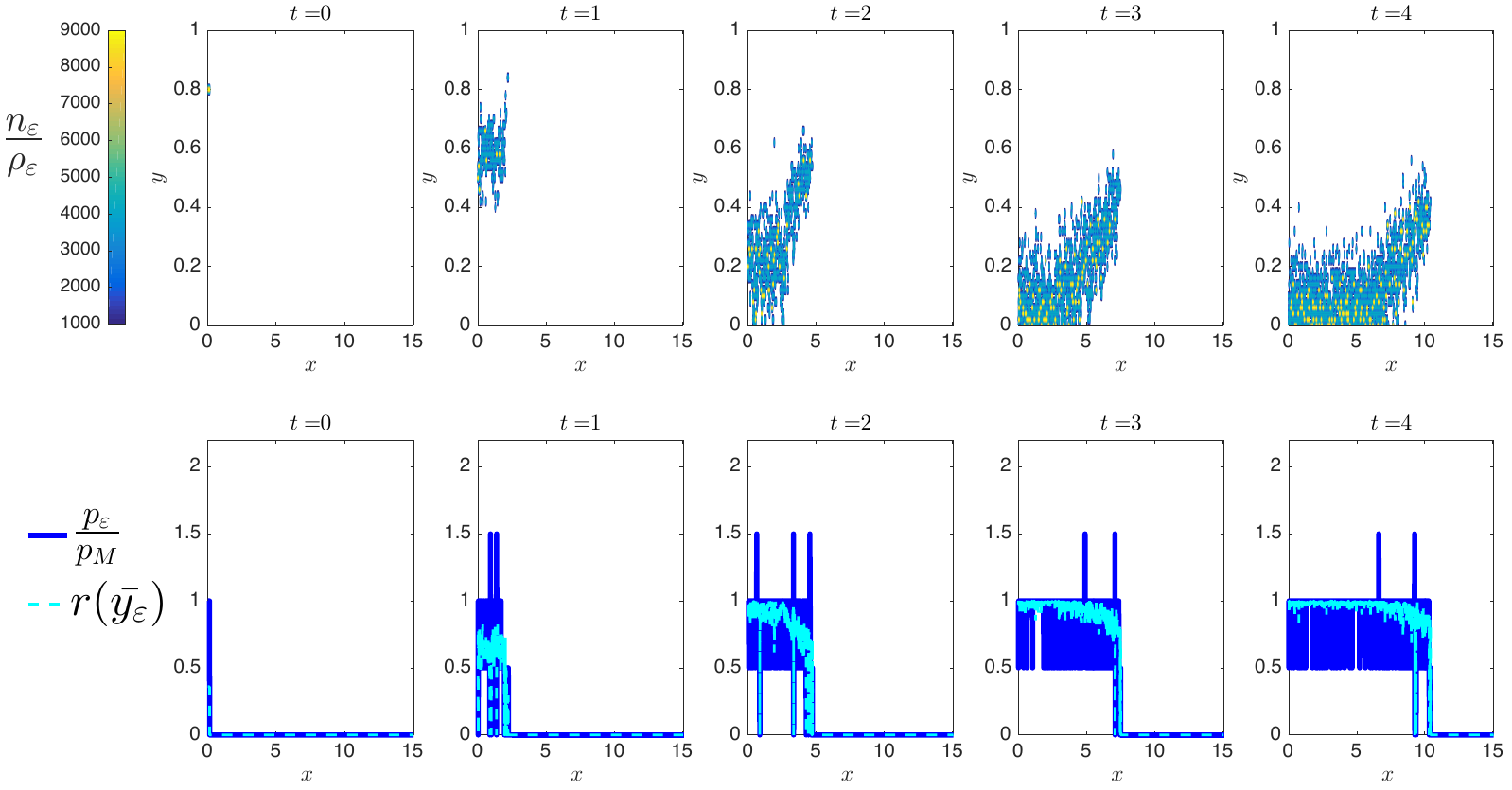}
\caption{Numerical results of {{a single simulation of the}} IB model with $A_0=0.27$ and $\bar{y}^0=0.8$ in the initial cell distribution~\eqref{def:icIB} {{-- \emph{i.e.}}} 1 out of the 10 simulations that are used to produce the average results displayed in the right-column, bottom-line panel of Figure~\ref{fig:test_P4a}. Plots display the scaled cell population density $n_\varepsilon/\rho_\varepsilon$ (top panels) and the scaled cell pressure $p_\varepsilon/ p_M$ (bottom panels, solid blue lines) at five successive time instants (\emph{i.e.} $t = 0$, $t = 1$, $t = 2$, $t = 3$ and $t = 4$). The dashed cyan lines in the bottom panels highlight the corresponding values of $r(\bar{y}_\varepsilon)$, with $\bar{y}_\varepsilon$ defined via~\eqref{def:yeps}. In this simulation, the cell population does not go extinct.}  
\label{fig:test_P5}
\end{figure}

\begin{figure}[tbhp]
\centering
\includegraphics[height=0.5\textwidth]{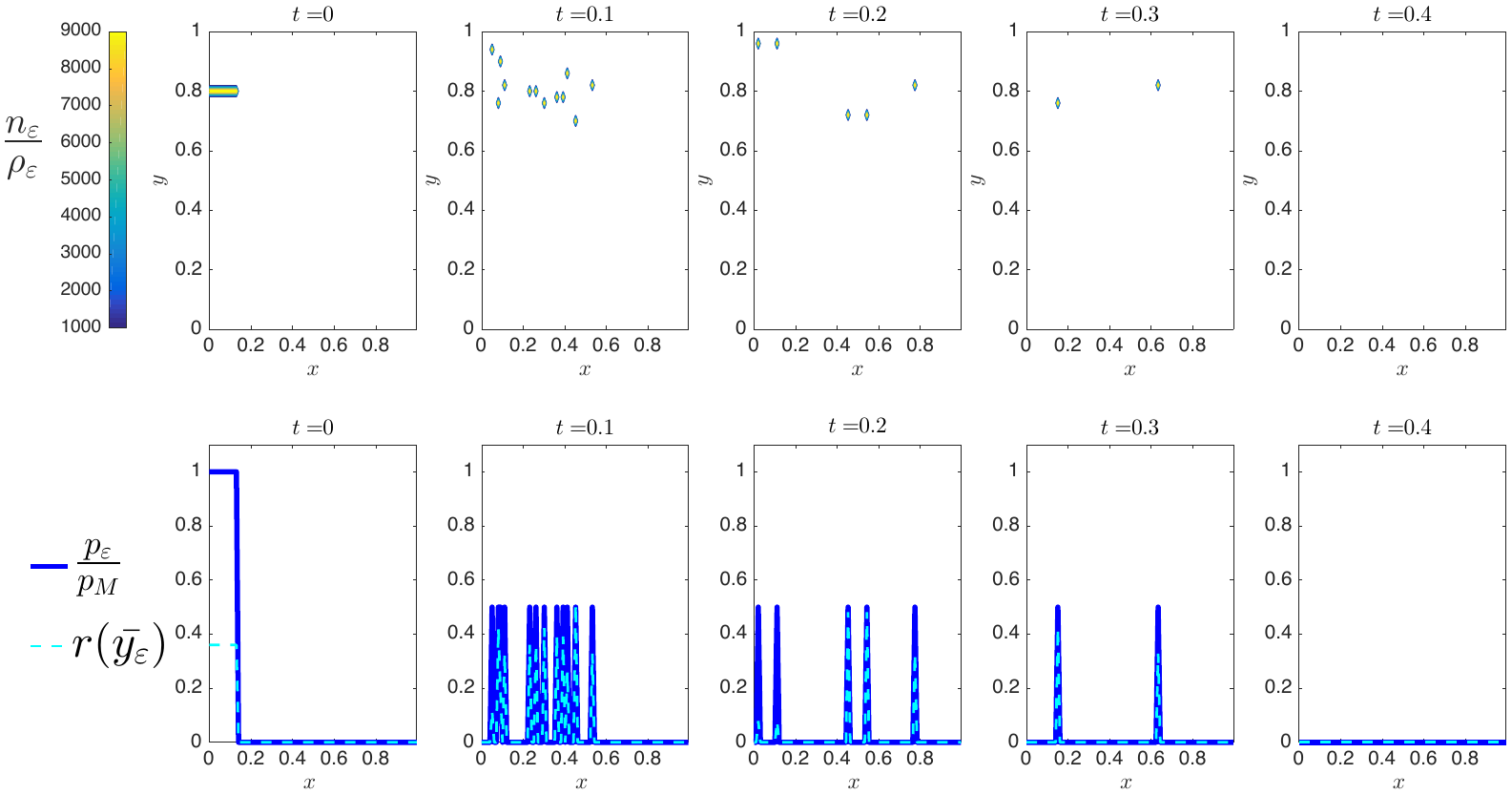}
\caption{Numerical results of {{a single simulation of the}} IB model with $A_0=0.27$ and $\bar{y}^0=0.8$ in the initial cell distribution~\eqref{def:icIB} {{-- \emph{i.e.}}} 1 out of the 10 simulations that are used to produce the average results displayed in the right-column, bottom-line panel of Figure~\ref{fig:test_P4a}. Plots display the scaled cell population density $n_\varepsilon/\rho_\varepsilon$ (top panels) and the scaled cell pressure $p_\varepsilon/ p_M$ (bottom panels, solid blue lines) at five successive time instants (\emph{i.e.} $t = 0$, $t = 0.1$, $t = 0.2$, $t = 0.3$ and $t = 0.4$). The dashed cyan lines in the bottom panels highlight the corresponding values of $r(\bar{y}_\varepsilon)$, with $\bar{y}_\varepsilon$ defined via~\eqref{def:yeps}. In this simulation, the cell population goes extinct rapidly.}  
\label{fig:test_P6}
\end{figure}

\newpage
{{In line with our expectations, this is due to the fact that stronger stochastic effects associated with small population levels in the initial phase of cell dynamics create the potential for population extinction to occur in some simulations of the IB model -- i.e. under the exact same parameter setting, we can observe extinction or survival of the population in the IB model due to demographic stochasticity (cf. the single simulation results displayed in Figures~\ref{fig:test_P5} and~\ref{fig:test_P6}). On the other hand,  the cell population will always persist according to the PDE model.  This ultimately results in discrepancies between the average behaviour of the IB model and the behaviour of the PDE model (cf. the plots in the bottom-line panels of Figures~\ref{fig:test_P4a} and~\ref{fig:test_P4b}).}}

\newpage
%%%%%%%%%%%%%%%%%%%%%%%%%%%%%%%%%%%%%%%%%
%% Discussion %% 
%%%%%%%%%%%%%%%%%%%%%%%%%%%%%%%%%%%%%%%%%
\section{Conclusions and research perspectives}
\label{sec:conclusion}
We developed an IB model for the dynamics of phenotypically heterogeneous growing cell populations, which captures intercellular variability in cell proliferation and migration rates. We concentrated on a proliferation-migration tradeoff scenario, where the cell phenotypes span a spectrum of states from minimally-mobile but highly-proliferative to highly-mobile but minimally-proliferative. In the context of cancer invasion, such a tradeoff is the tenet of the ``go-or-grow'' hypothesis, which was conceived following observations of glioma cell behaviour~\cite{giese1996dichotomy} and has stimulated much empirical and theoretical research -- see, for instance, \cite{corcoran2003,gallaher2019impact,giese1996dichotomy,hatzikirou2012go,hoek2008vivo,pham2012density,stepien2018,vittadello2020,zhigun2018} and references therein.

We reported on the results of numerical simulations of the IB model which illustrate how proliferation-migration tradeoffs shaping the evolutionary dynamics of single cells can lead, at the population level, to the generation of travelling waves whereby phenotypes are structured across the support of the wave, with highly-mobile cells being found at the invasive front and more-proliferative cells dominating at the rear. Similar patterns of cell population growth have been observed in gliomas, where cells within the interior of the tumour exhibit higher proliferation and lower migration rates, while cells on the tumour border are instead characterised by lower proliferation and higher migration rates~\cite{dhruv2013reciprocal,giese2003cost,giese1996dichotomy,wang2012ephb2,xie2014targeting}.

We formally derived the deterministic continuum counterpart of the IB model, which comprises a non-local PDE for the cell population density function, and carried out a comparative study between numerical simulations of the IB model and both numerical solutions and formal travelling-wave analysis of the PDE model. We demonstrated that there is an excellent quantitative agreement between the results of numerical simulations of the IB model and the results of numerical simulations and travelling-wave analysis of the corresponding PDE model, when sufficiently large cell numbers are considered. This testifies to the robustness of the results of numerical simulations of the IB model presented here (see Remark~\ref{rem:robu}). 

{{In general, agreement between IB models and their continuum counterparts arises in regions of the model parameter space that correspond to sufficiently large cell numbers~\cite{simpson2022reliable,nardini2021learning,simpson2014distinguishing}, while discrepancies may arise when the number of cells becomes low -- \emph{e.g.} if the rate of cell death is sufficiently large~\cite{johnston2020predicting} -- leading to possible extinction of the population in the IB model. We have provided numerical evidence of situations such as these in which the predictions of the two models can differ due to demographic stochasticity, which cannot be captured by the PDE model. This indicates the importance of integrating individual-based and continuum approaches when modelling the growth of phenotypically heterogeneous cell populations. }}

Although in this work we focused on a one-dimensional spatial domain scenario, the IB model presented here, and the formal limiting procedure to derive the corresponding continuum model, could easily be adapted to higher spatial dimensions. Furthermore, while we represented the spatial domain of the IB model as a regular lattice, it would certainly be interesting to generalise the underlying modelling approach, as well as the formal method to derive the continuum counterpart of the model, to cases where cells are distributed over irregular lattices and also to cases where off-latice representations of the spatial domain are adopted. The present IB model could also be extended further to include the effects of chemical species (\emph{e.g.} nutrients, growth factors, chemoattractants, chemorepulsants) and how the cells interact with and respond to these chemicals. {{To include and implement chemical species in the current model, we could use a hybrid modelling approach whereby the probabilistic rules governing the dynamics of single cells would be coupled with balance equations for the chemical concentrations. Hybrid modelling approaches of this type have been utilised in the context of modelling various aspects of cancer growth and development -- see, for instance, ~\cite{lowengrub2009nonlinear,powathil2015systems,rejniak2011hybrid,jafari2021multiscale,jafari2022multiscale,bubba2020discrete}.}}

The generality of our assumptions makes the IB modelling framework presented here applicable to a broad range of biological processes that are driven by the growth of phenotypically heterogeneous cell populations, including tumour invasion and tissue remodelling and repair. It would thus be interesting to focus on particular cellular systems, and consequently define specific models, {{which could then be more accurately parameterised using precise biological data}}. This would offer the opportunity to dissect out the role played by different spatiotemporal evolutionary processes at the single-cell level in the formation of complex spatial patterns of population growth. 

\section*{Acknowledgments}
T.L. gratefully acknowledges support from the MIUR grant ``Dipartimenti di Eccellenza 2018-2022'' (Project no. E11G18000350001). F.R.M. gratefully acknowledges support from the RSE Saltire Early Career Fellowship `Multiscale mathematical modelling of spatial eco-evolutionary cancer dynamics' (Fellowship No. 1879).

\section*{Conflict of interest}
The authors declare no competing interests.

%%%%%%%%%%%%%%%%%%%%%%%%%%%%%%%%%%%%%%%%%
%% Bibliography %% 
%%%%%%%%%%%%%%%%%%%%%%%%%%%%%%%%%%%%%%%%%

\bibliographystyle{plain}
\bibliography{references_final}

\newpage
%%%%%%%%%%%%%%%%%%%%%%%%%%%%%%%%%%%%%%%%%
%% Appendices %% 
%%%%%%%%%%%%%%%%%%%%%%%%%%%%%%%%%%%%%%%%%

\appendix 
\section*{Appendices}
\section{Formal derivation of the continuum model}
\label{app:derivation}
Building on the methods that we previously employed in~\cite{ardavseva2020comparative,bubba2020discrete,chaplain2020bridging,macfarlane2020hybrid,stace2020discrete}, here we show that the non-local PDE (3.2) can be formally derived as the appropriate continuum limit of the IB model developed in this paper.\\ 

In the case where, between time-steps $k$ and $k+1$, each cell in phenotypic state $y_j \in (0,Y)$ at position $x_i \in \mathbb{R}$ can first move, next undergo phenotypic changes and then die or divide according to the rules described in Section 2, the principle of mass balance gives the following difference equation
\begin{eqnarray}
\label{Der1}
n^{k+1}_{i,j}&=&n^{k}_{i+1,j+1}\left\{ \frac{\lambda}{2}\left[ 1+\tau R(y_j, p^{k}_{i})\right]\frac{\nu\mu(y_{j})}{2p_M} \left(p^{k}_{i+1}-p^{k}_{i}\right)_+\right\}\\
&&+n^{k}_{i-1,j+1}\left\{ \frac{\lambda}{2}\left[ 1+\tau R(y_j, p^{k}_{i})\right]\frac{\nu\mu(y_{j})}{2p_M} \left(p^{k}_{i-1}-p^{k}_{i}\right)_+\right\}\nonumber\\
&&+n^{k}_{i+1,j-1}\left\{ \frac{\lambda}{2}\left[ 1+\tau R(y_j, p^{k}_{i})\right]\frac{\nu\mu(y_{j})}{2p_M} \left(p^{k}_{i+1}-p^{k}_{i}\right)_+\right\}\nonumber\\
&&+n^{k}_{i-1,j-1}\left\{ \frac{\lambda}{2}\left[ 1+\tau R(y_j, p^{k}_{i})\right]\frac{\nu\mu(y_{j})}{2p_M} \left(p^{k}_{i-1}-p^{k}_{i}\right)_+\right\}\nonumber\\
&&+n^{k}_{i,j+1}\left\{ \frac{\lambda}{2}\left[ 1+\tau R(y_j, p^{k}_{i})\right]\left[1-\frac{\nu\mu(y_{j})}{2p_M}\left[ \left(p^{k}_{i}-p^{k}_{i+1}\right)_+ + \left(p^{k}_{i}-p^{k}_{i-1}\right)_+\right]\right] \right\}\nonumber\\
&&+n^{k}_{i,j-1}\left\{ \frac{\lambda}{2}\left[ 1+\tau R(y_j, p^{k}_{i})\right]\left[1-\frac{\nu\mu(y_{j})}{2p_M}\left[ \left(p^{k}_{i}-p^{k}_{i+1}\right)_+ + \left(p^{k}_{i}-p^{k}_{i-1}\right)_+\right]\right] \right\}\nonumber\\
&&+n^{k}_{i+1,j}\left\{ (1-\lambda)\left[ 1+\tau R(y_j, p^{k}_{i})\right]\frac{\nu\mu(y_{j})}{2p_M} \left(p^{k}_{i+1}-p^{k}_{i}\right)_+\right\}\nonumber\\
&&+n^{k}_{i-1,j}\left\{ (1-\lambda)\left[ 1+\tau R(y_j, p^{k}_{i})\right]\frac{\nu\mu(y_{j})}{2p_M} \left(p^{k}_{i-1}-p^{k}_{i}\right)_+\right\}\nonumber\\
&&+n^{k}_{i,j}\left\{ (1-\lambda)\left[ 1+\tau R(y_j, p^{k}_{i})\right] \left[1-\frac{\nu\mu(y_{j})}{2p_M}\left[\left(p^{k}_{i}-p^{k}_{i+1}\right)_++\left(p^{k}_{i}-p^{k}_{i-1}\right)_+ \right]\right]\nonumber\right\}.
\end{eqnarray}
Using the fact that for $\tau$, $\chi$ and $\eta$ sufficiently small the following relations hold
 \[
n^{k}_{i,j}\approx n(t,x,y),\quad n^{k+1}_{i,j}\approx n(t+\tau,x,y),\quad n^{k}_{i\pm1,j}\approx n(t,x\pm\chi,y),\quad n^{k}_{i,j\pm1}\approx n(t,x,y\pm \eta)
\]
 \[
 \rho^{k}_{i} \approx \rho(t,x) := \int_0^Y n(t,x,y) \, {\rm d}y, \quad  p^{k}_{i} \approx p(t,x) = \Pi[\rho](t,x), \quad p^{k}_{i\pm1} \approx p(t,x\pm\chi) = \Pi[\rho](t,x\pm\chi),
\]
equation~\eqref{Der1} can be formally rewritten in the approximate form
\begin{eqnarray}
\label{Der2}
n(t+\tau,x,y)&=&n(t,x+\chi,y+\eta)\left\{ \frac{\lambda}{2}\left[ 1+\tau R(y,p)\right]\frac{\nu\mu(y)}{2p_M} \left(p(t,x+\chi)-p\right)_+\right\}\\
&&+n(t,x-\chi,y+\eta)\left\{ \frac{\lambda}{2}\left[ 1+\tau R(y,p)\right]\frac{\nu\mu(y)}{2p_M} \left(p(t,x-\chi)-p\right)_+\right\}\nonumber\\
&&+n(t,x+\chi,y-\eta)\left\{ \frac{\lambda}{2}\left[ 1+\tau R(y,p)\right]\frac{\nu\mu(y)}{2p_M} \left(p(t,x+\chi)-p\right)_+\right\}\nonumber\\
&&+n(t,x-\chi,y-\eta)\left\{ \frac{\lambda}{2}\left[ 1+\tau R(y,p)\right]\frac{\nu\mu(y)}{2p_M} \left(p(t,x-\chi)-p\right)_+\right\}\nonumber\\
&&+n(t,x,y+\eta)\left\{ \frac{\lambda}{2}\left[ 1+\tau R(y,p)\right]\left[1-\frac{\nu\mu(y)}{2p_M}\left[ \left(p-p(t,x+\chi)\right)_+ + \left(p-p(t,x-\chi)\right)_+\right]\right] \right\}\nonumber\\
&&+n(t,x,y-\eta)\left\{ \frac{\lambda}{2}\left[ 1+\tau R(y,p)\right]\left[1-\frac{\nu\mu(y)}{2p_M}\left[ \left(p-p(t,x+\chi)\right)_+ + \left(p-p(t,x-\chi)\right)_+\right]\right] \right\}\nonumber\\
&&+n(t,x+\chi,y)\left\{ (1-\lambda)\left[ 1+\tau R(y,p)\right]\frac{\nu\mu(y)}{2p_M} \left(p(t,x+\chi)-p\right)_+\right\}\nonumber\\
&&+n(t,x-\chi,y)\left\{ (1-\lambda)\left[ 1+\tau R(y,p)\right]\frac{\nu\mu(y)}{2p_M} \left(p(t,x-\chi)-p\right)_+\right\}\nonumber\\
&&+n\left\{ (1-\lambda)\left[ 1+\tau R(y,p)\right] \left[1-\frac{\nu\mu(y)}{2p_M}\left(p-p(t,x+\chi)\right)_+-\frac{\nu\mu(y)}{2p_M}\left(p-p(t,x-\chi)\right)_+ \right]\right\},\nonumber
\end{eqnarray}
where $n \equiv n(t,x,y)$ and $p \equiv p(t,x)$. If the function $n(t,x,y)$ is twice continuously differentiable with respect to the variables $y$ and $x$, for $\eta$ and $\chi$ sufficiently small we can then use the Taylor expansions
\[
 n(t,x,y\pm\eta) = n\pm\eta\frac{\partial n}{\partial y}+\frac{\eta^2}{2}\frac{\partial^2 n}{\partial y^2}+\text{h.o.t.} \, , \quad n(t,x\pm\chi,y)= n\pm\chi\frac{\partial n}{\partial x}+\frac{\chi^2}{2}\frac{\partial^2 n}{\partial x^2}+\text{h.o.t.} \, ,
\]
\[
n(t,x+\chi,y\pm\eta)= n+\chi\frac{\partial n}{\partial x} \pm\eta\frac{\partial n}{\partial y}+\frac{\chi^2}{2}\frac{\partial^2 n}{\partial x^2}+\frac{\eta^2}{2}\frac{\partial^2 n}{\partial y^2} \pm \chi\eta \frac{\partial^2 n}{\partial x\partial y}+\text{h.o.t.}
\]
and
\[
n(t,x-\chi,y\pm\eta)= n-\chi\frac{\partial n}{\partial x} \pm\eta\frac{\partial n}{\partial y}+\frac{\chi^2}{2}\frac{\partial^2 n}{\partial x^2}+\frac{\eta^2}{2}\frac{\partial^2 n}{\partial y^2} \mp \chi\eta \frac{\partial^2 n}{\partial x\partial y}+\text{h.o.t.} \, ,
\]
which allow us to rewrite equation~\eqref{Der2} as

\begin{eqnarray}
\label{Der3}
n(t+\tau,x,y)&=&n\left\{ \frac{\lambda}{2}\left[ 1+\tau R(y,p)\right]\frac{\nu\mu(y)}{2p_M} \left(p(t,x+\chi)-p\right)_+\right\}\\
&&+\chi\frac{\partial n}{\partial x}\left\{ \frac{\lambda}{2}\left[ 1+\tau R(y,p)\right]\frac{\nu\mu(y)}{2p_M} \left(p(t,x+\chi)-p\right)_+\right\}\nonumber\\
&&+\eta\frac{\partial n}{\partial y}\left\{ \frac{\lambda}{2}\left[ 1+\tau R(y,p)\right]\frac{\nu\mu(y)}{2p_M} \left(p(t,x+\chi)-p\right)_+\right\}\nonumber\\
&&+\frac{\chi^2}{2}\frac{\partial^2 n}{\partial x^2}\left\{ \frac{\lambda}{2}\left[ 1+\tau R(y,p)\right]\frac{\nu\mu(y)}{2p_M} \left(p(t,x+\chi)-p\right)_+\right\}\nonumber\\
&&+\frac{\eta^2}{2}\frac{\partial^2 n}{\partial y^2}\left\{ \frac{\lambda}{2}\left[ 1+\tau R(y,p)\right]\frac{\nu\mu(y)}{2p_M} \left(p(t,x+\chi)-p\right)_+\right\}\nonumber\\
&&+\chi\eta\frac{\partial^2 n}{\partial x \partial y}\left\{ \frac{\lambda}{2}\left[ 1+\tau R(y,p)\right]\frac{\nu\mu(y)}{2p_M} \left(p(t,x+\chi)-p\right)_+\right\}\nonumber\\
%%%
&&+n\left\{ \frac{\lambda}{2}\left[ 1+\tau R(y,p)\right]\frac{\nu\mu(y)}{2p_M} \left(p(t,x-\chi)-p\right)_+\right\}\nonumber\\
&&-\chi \frac{\partial n}{\partial x}\left\{ \frac{\lambda}{2}\left[ 1+\tau R(y,p)\right]\frac{\nu\mu(y)}{2p_M} \left(p(t,x-\chi)-p\right)_+\right\}\nonumber\\
&&+\eta\frac{\partial n}{\partial y}\left\{ \frac{\lambda}{2}\left[ 1+\tau R(y,p)\right]\frac{\nu\mu(y)}{2p_M} \left(p(t,x-\chi)-p\right)_+\right\}\nonumber\\
&&+\frac{\chi^2}{2}\frac{\partial^2 n}{\partial x^2}\left\{ \frac{\lambda}{2}\left[ 1+\tau R(y,p)\right]\frac{\nu\mu(y)}{2p_M} \left(p(t,x-\chi)-p\right)_+\right\}\nonumber\\
&&+\frac{\eta^2}{2}\frac{\partial^2 n}{\partial y^2}\left\{ \frac{\lambda}{2}\left[ 1+\tau R(y,p)\right]\frac{\nu\mu(y)}{2p_M} \left(p(t,x-\chi)-p\right)_+\right\}\nonumber\\
&&-\chi\eta\frac{\partial^2 n}{\partial x\partial y}\left\{ \frac{\lambda}{2}\left[ 1+\tau R(y,p)\right]\frac{\nu\mu(y)}{2p_M} \left(p(t,x-\chi)-p\right)_+\right\}\nonumber\\
%%%
&&+n\left\{ \frac{\lambda}{2}\left[ 1+\tau R(y,p)\right]\frac{\nu\mu(y)}{2p_M} \left(p(t,x+\chi)-p\right)_+\right\}\nonumber\\
&&+\chi \frac{\partial n}{\partial x}\left\{ \frac{\lambda}{2}\left[ 1+\tau R(y,p)\right]\frac{\nu\mu(y)}{2p_M} \left(p(t,x+\chi)-p\right)_+\right\}\nonumber\\
&&-\eta \frac{\partial n}{\partial y}\left\{ \frac{\lambda}{2}\left[ 1+\tau R(y,p)\right]\frac{\nu\mu(y)}{2p_M} \left(p(t,x+\chi)-p\right)_+\right\}\nonumber\\
&&+\frac{\chi^2}{2}\frac{\partial^2 n}{\partial x^2}\left\{ \frac{\lambda}{2}\left[ 1+\tau R(y,p)\right]\frac{\nu\mu(y)}{2p_M} \left(p(t,x+\chi)-p\right)_+\right\}\nonumber\\
&&+\frac{\eta^2}{2}\frac{\partial^2 n}{\partial y^2}\left\{ \frac{\lambda}{2}\left[ 1+\tau R(y,p)\right]\frac{\nu\mu(y)}{2p_M} \left(p(t,x+\chi)-p\right)_+\right\}\nonumber\\
&&-\chi\eta\frac{\partial^2 n}{\partial x\partial y}\left\{ \frac{\lambda}{2}\left[ 1+\tau R(y,p)\right]\frac{\nu\mu(y)}{2p_M} \left(p(t,x+\chi)-p\right)_+\right\}\nonumber\\
%%%
&&+n\left\{ \frac{\lambda}{2}\left[ 1+\tau R(y,p)\right]\frac{\nu\mu(y)}{2p_M} \left(p(t,x-\chi)-p\right)_+\right\}\nonumber\\
&&-\chi \frac{\partial n}{\partial x}\left\{ \frac{\lambda}{2}\left[ 1+\tau R(y,p)\right]\frac{\nu\mu(y)}{2p_M} \left(p(t,x-\chi)-p\right)_+\right\}\nonumber\\
&&-\eta \frac{\partial n}{\partial y}\left\{ \frac{\lambda}{2}\left[ 1+\tau R(y,p)\right]\frac{\nu\mu(y)}{2p_M} \left(p(t,x-\chi)-p\right)_+\right\}\nonumber\\
&&+\frac{\chi^2}{2}\frac{\partial^2 n}{\partial x^2}\left\{ \frac{\lambda}{2}\left[ 1+\tau R(y,p)\right]\frac{\nu\mu(y)}{2p_M} \left(p(t,x-\chi)-p\right)_+\right\}\nonumber\\
&&+\frac{\eta^2}{2}\frac{\partial^2 n}{\partial y^2}\left\{ \frac{\lambda}{2}\left[ 1+\tau R(y,p)\right]\frac{\nu\mu(y)}{2p_M} \left(p(t,x-\chi)-p\right)_+\right\}\nonumber\\
&&+\chi\eta\frac{\partial^2 n}{\partial x\partial y}\left\{ \frac{\lambda}{2}\left[ 1+\tau R(y,p)\right]\frac{\nu\mu(y)}{2p_M} \left(p(t,x-\chi)-p\right)_+\right\}\nonumber\\
%%%
&&+n\left\{ \frac{\lambda}{2}\left[ 1+\tau R(y,p)\right]\left[1-\frac{\nu\mu(y)}{2p_M}\left[ \left(p-p(t,x+\chi)\right)_+ + \left(p-p(t,x-\chi)\right)_+\right]\right] \right\}\nonumber\\
&&+\eta \frac{\partial n}{\partial y}\left\{ \frac{\lambda}{2}\left[ 1+\tau R(y,p)\right]\left[1-\frac{\nu\mu(y)}{2p_M}\left[ \left(p-p(t,x+\chi)\right)_+ + \left(p-p(t,x-\chi)\right)_+\right]\right] \right\}\nonumber\\
&&+\frac{\eta^2}{2}\frac{\partial^2 n}{\partial y^2}\left\{ \frac{\lambda}{2}\left[ 1+\tau R(y,p)\right]\left[1-\frac{\nu\mu(y)}{2p_M}\left[ \left(p-p(t,x+\chi)\right)_+ + \left(p-p(t,x-\chi)\right)_+\right]\right] \right\}\nonumber\\
%%%
&&+n\left\{ \frac{\lambda}{2}\left[ 1+\tau R(y,p)\right]\left[1-\frac{\nu\mu(y)}{2p_M}\left[ \left(p-p(t,x+\chi)\right)_+ + \left(p-p(t,x-\chi)\right)_+\right]\right] \right\}\nonumber\\
&&-\eta  \frac{\partial n}{\partial y}\left\{ \frac{\lambda}{2}\left[ 1+\tau R(y,p)\right]\left[1-\frac{\nu\mu(y)}{2p_M}\left[ \left(p-p(t,x+\chi)\right)_+ + \left(p-p(t,x-\chi)\right)_+\right]\right] \right\}\nonumber\\
&&+\frac{\eta^2}{2}\frac{\partial^2 n}{\partial y^2}\left\{ \frac{\lambda}{2}\left[ 1+\tau R(y,p)\right]\left[1-\frac{\nu\mu(y)}{2p_M}\left[ \left(p-p(t,x+\chi)\right)_+ + \left(p-p(t,x-\chi)\right)_+\right]\right] \right\}\nonumber\\
%%%
&&+n\left\{ (1-\lambda)\left[ 1+\tau R(y,p)\right]\frac{\nu\mu(y)}{2p_M} \left(p(t,x+\chi)-p\right)_+\right\}\nonumber\\
&&+\chi  \frac{\partial n}{\partial x} \left\{ (1-\lambda)\left[ 1+\tau R(y,p)\right]\frac{\nu\mu(y)}{2p_M} \left(p(t,x+\chi)-p\right)_+\right\}\nonumber\\
&&+\frac{\chi^2}{2} \frac{\partial^2 n}{\partial x^2} \left\{ (1-\lambda)\left[ 1+\tau R(y,p)\right]\frac{\nu\mu(y)}{2p_M} \left(p(t,x+\chi)-p\right)_+\right\}\nonumber\\
%%%
&&+n\left\{ (1-\lambda)\left[ 1+\tau R(y,p)\right]\frac{\nu\mu(y)}{2p_M} \left(p(t,x-\chi)-p\right)_+\right\}\nonumber\\
&&-\chi  \frac{\partial n}{\partial x}\left\{ (1-\lambda)\left[ 1+\tau R(y,p)\right]\frac{\nu\mu(y)}{2p_M} \left(p(t,x-\chi)-p\right)_+\right\}\nonumber\\
&&+\frac{\chi^2}{2} \frac{\partial^2 n}{\partial x^2}\left\{ (1-\lambda)\left[ 1+\tau R(y,p)\right]\frac{\nu\mu(y)}{2p_M} \left(p(t,x-\chi)-p\right)_+\right\}\nonumber\\
%%%
&&+n\left\{ (1-\lambda)\left[ 1+\tau R(y,p)\right] \left[1-\frac{\nu\mu(y)}{2p_M}\left(p-p(t,x+\chi)\right)_+-\frac{\nu\mu(y)}{2p_M}\left(p-p(t,x-\chi)\right)_+ \right]\right\} + \text{h.o.t.} \, .\nonumber
\end{eqnarray}

Collecting terms that contain the same derivative of $n$ we can further simplify equation~\eqref{Der3} to obtain
\begin{eqnarray}
\label{Der4}
n(t+\tau,x,y)&=&n \left[ 1+\tau R(y,p)\right] \\
&&+n\left\{\left[ 1+\tau R(y,p)\right]\frac{\nu\mu(y)}{2p_M}\left[ \left(p(t,x+\chi)-p\right)_++\left(p(t,x-\chi)-p\right)_+\right]\right\}\nonumber\\
&&-n\left\{ \left[ 1+\tau R(y,p)\right]\frac{\nu\mu(y)}{2p_M} \left[\left(p-p(t,x+\chi)\right)_++\left(p-p(t,x-\chi)\right)_+ \right]\right\}\nonumber\\
%%%%%%%
&&+\chi  \frac{\partial n}{\partial x} \left\{ \left[ 1+\tau R(y,p)\right]\frac{\nu\mu(y)}{2p_M} \left[\left(p(t,x+\chi)-p\right)_+-\left(p(t,x-\chi)-p\right)_+\right]\right\}\nonumber\\
%%%%%%%
&&+\frac{\chi^2}{2} \frac{\partial^2 n}{\partial x^2} \left\{ \left[ 1+\tau R(y,p)\right]\frac{\nu\mu(y)}{2p_M} \left[\left(p(t,x+\chi)-p\right)_++\left(p(t,x-\chi)-p\right)_+\right]\right\}\nonumber\\
%%%%%%%
&&+\frac{\eta^2}{2}\frac{\partial^2 n}{\partial y^2}\left\{ \lambda\left[ 1+\tau R(y,p)\right]\frac{\nu\mu(y)}{2p_M} \left[\left(p(t,x+\chi)-p\right)_++ \left(p(t,x-\chi)-p\right)_+\right]\right\}\nonumber\\
&&-\frac{\eta^2}{2}\frac{\partial^2 n}{\partial y^2}\left\{ \lambda\left[ 1+\tau R(y,p)\right]\frac{\nu\mu(y)}{2p_M}\left[ \left(p-p(t,x+\chi)\right)_+ + \left(p-p(t,x-\chi)\right)_+\right] \right\}\nonumber\\
&&+\frac{\lambda \eta^2 }{2}\frac{\partial^2 n}{\partial y^2}\left[ 1+\tau R(y,p)\right] + \text{h.o.t.} \, .\nonumber
\end{eqnarray}
Rewriting the above equation by using the fact that 
\begin{eqnarray*}
&&\left[ \left(p(t,x+\chi)-p\right)_++\left(p(t,x-\chi)-p\right)_+\right] \\
&& \qquad \qquad - \left[\left(p-p(t,x+\chi)\right)_++\left(p-p(t,x-\chi)\right)_+ \right] = p(t,x+\chi)+p(t,x-\chi)-2 \, p,
\end{eqnarray*}
dividing both sides of the resulting equation by $\tau$, rearranging terms and then multiplying and dividing the terms on the right-hand side by either $\chi^2$ or $\chi$ we find 
\begin{eqnarray*}
%\label{Der5}
\frac{n(t+\tau,x,y)-n}{\tau} &=&  R(y,p) \, n +\frac{\lambda \eta^2 }{2\tau}\frac{\partial^2 n}{\partial y^2}\left[ 1+\tau R(y,p)\right] \\
&&+\frac{\nu \chi^2}{2\tau} \, n\left\{\left[ 1+\tau R(y,p)\right]\frac{\mu(y)}{p_M}\left[\dfrac{p(t,x+\chi)+p(t,x-\chi)-2 \, p}{\chi^2}\right]\right\}\nonumber\\
%%%%%%%
&&+\frac{\nu \chi^2}{2\tau} \, \frac{\partial n}{\partial x} \left\{ \left[ 1+\tau R(y,p)\right]\frac{\mu(y)}{p_M} \left[\left(\dfrac{p(t,x+\chi)-p}{\chi}\right)_+-\left(\dfrac{p(t,x-\chi)-p}{\chi}\right)_+\right]\right\}\nonumber\\
%%%%%%%
&&+\frac{\chi}{2} \, \frac{\nu \chi^2}{2\tau} \frac{\partial^2 n}{\partial x^2} \left\{ \left[ 1+\tau R(y,p)\right]\frac{\mu(y)}{p_M} \left[\left(\dfrac{p(t,x+\chi)-p}{\chi}\right)_++\left(\dfrac{p(t,x-\chi)-p}{\chi}\right)_+\right]\right\}\nonumber\\
%%%%%%%
&&+\frac{\eta^2}{2} \, \frac{\nu \chi^2}{2 \tau}\frac{\partial^2 n}{\partial y^2}\left\{ \lambda\left[ 1+\tau R(y,p)\right]\frac{\mu(y)}{p_M} \left[\dfrac{p(t,x+\chi)+p(t,x-\chi)-2 \, p}{\chi^2}\right]\right\}\nonumber\\
&& + \, \text{h.o.t.} \, . \nonumber
\end{eqnarray*}
If the function $n(t,x,y)$ is also continuously differentiable with respect to the variable $t$ and the function $p(t,x)$ is twice continuously differentiable with respect to the variable $x$,  letting $\tau\rightarrow 0$, $\chi\rightarrow 0$ and $\eta\rightarrow0$ in such a way that conditions (3.1) are met, from the latter equation we formally obtain 
$$
\frac{\partial n}{\partial t}=R(y,p) \, n + \beta \frac{\partial^2 n}{\partial y^2} + \alpha \, \dfrac{\mu(y)}{p_M} \left\{n\frac{\partial^2 p}{\partial x^2}+  \frac{\partial n}{\partial x} \left[\left(\frac{\partial p}{\partial x}\right)_+ - \left(-\frac{\partial p}{\partial x}\right)_+\right]\right\}.
%, \quad (x,y) \in \mathbb{R} \times (0,Y),
$$
Hence, using the definition $\hat{\mu}(y) := \dfrac{\mu(y)}{p_M}$ along with the fact that $\displaystyle{\left(\frac{\partial p}{\partial x}\right)_+ - \left(-\frac{\partial p}{\partial x}\right)_+ = \frac{\partial p}{\partial x}}$, and recalling that $(x,y) \in \mathbb{R} \times (0,Y)$, we find the following non-local PDE for the cell population density function $n(t,x,y)$ 
$$
\frac{\partial n}{\partial t}=R(y,p) \, n + \beta \frac{\partial^2 n}{\partial y^2} + \alpha \, \hat{\mu}(y) \left[n\frac{\partial^2 p}{\partial x^2}+  \frac{\partial n}{\partial x}\frac{\partial p}{\partial x}\right], \quad (x,y) \in \mathbb{R} \times (0,Y),
$$
which can easily be rewritten as the non-local PDE (3.2). Finally, zero-Neumann (\emph{i.e.} no-flux) boundary conditions at $y=0$ and $y=Y$ follow from the fact that the attempted phenotypic variation of a cell is aborted if it requires moving into a phenotypic state that does not belong to the interval $[0,Y]$.

\section{Formal travelling-wave analysis for $\varepsilon \to 0$}
\label{TWanalysis}
Adopting a method analogous to those that we used~\cite{lorenzi2021,lorenzi2022trade}, which build on the Hamilton-Jacobi approach developed in~\cite{barles2009concentration,diekmann2005dynamics,lorz2011dirac,perthame2006transport,perthame2008dirac}, we make the real phase WKB ansatz~\cite{barles1989wavefront,evans1989pde,fleming1986pde}
\begin{equation} \label{WKB}
n_{\varepsilon}(t,x,y) = e^{\frac{u_{\varepsilon}(t,x,y)}{\varepsilon}},
\end{equation}
which gives
\[
\partial_t n_{\varepsilon} = \frac{\partial_t u_{\varepsilon}}{\varepsilon} n_{\varepsilon}, \quad \partial_x n_{\varepsilon} = \frac{\partial_x u_{\varepsilon}}{\varepsilon} n_{\varepsilon}, \quad \partial^2_{yy} n_{\varepsilon} = \left(\frac{1}{\varepsilon^2} \left(\partial_y u_{\varepsilon} \right)^2 + \frac{1}{\varepsilon} \partial^2_{yy} u_{\varepsilon} \right) n_{\varepsilon}.
\]
Substituting the above expressions into the non-local PDE (4.2) gives the following Hamilton-Jacobi equation for $u_{\varepsilon}(t,x,y)$ 
\begin{equation}
\label{eq:PDEue}
\partial_t u_{\varepsilon} - \hat{\mu}(y) \left(\partial_{x} u_{\varepsilon} \, \partial_{x} p_{\varepsilon} + \varepsilon \, \partial^2_{xx} p_{\varepsilon} \right) = R(y,p_{\varepsilon}) + \left(\partial_y u_{\varepsilon} \right)^2 + \varepsilon \, \partial^2_{yy} u_{\varepsilon}, \quad (x,y) \in \mathbb{R} \times(0,Y).
\end{equation}
Letting $\varepsilon \to 0$ in~\eqref{eq:PDEue} we formally obtain the following equation for the leading-order term $u(t,x,y)$ of the asymptotic expansion for $u_{\varepsilon}(t,x,y)$
\begin{equation}
\label{eq:PDEu}
\partial_t u - \hat{\mu}(y) \, \partial_{x} p \, \partial_{x} u = R(y,p) + \left(\partial_y u \right)^2, \quad (x,y) \in \mathbb{R} \times(0,Y),
\end{equation}
where $p(t,x)$ is the leading-order term of the asymptotic expansion for $p_{\varepsilon}(t,x)$. 

\paragraph{Constraint on $u$.}
Consider $x \in \mathbb{R}$ such that $\rho(t,x) >0$, that is, $x \in {\rm Supp}(\rho)$, and let $\bar{y}(t,x)$ be a non-degenerate maximum point of $u(t,x,y)$, that is, $\displaystyle{\bar{y}(t,x) \in \argmax_{y \in [0,Y]} u(t,x,y)}$ with $\partial^2_{yy} u(t,x,\bar{y})<0$. 
Letting $\varepsilon \to 0$ in~\eqref{WKB} formally gives the following constraint for all $t>0$
\begin{equation}
\label{eq:ubaryiszero}
u(t,x,\bar{y}(t,x)) = \max_{y \in [0,Y]} u(t,x,y) = 0, \quad x \in {\rm Supp}(\rho),
\end{equation}
which also implies that 
\begin{equation}
\label{eq:uybaryiszero}
\partial_y u(t,x,\bar{y}(t,x)) = 0 \quad \text{and} \quad \partial_x u(t,x,\bar{y}(t,x)) = 0, \quad x \in {\rm Supp}(\rho).
\end{equation}

\begin{remark}
When $n_{\varepsilon}(t,x,y)$ is in the form~\eqref{WKB}, if $u(t,x,y)$ is a strictly concave function of $y$ with maximum point $y=\bar{y}(t,x)$ then the constraint~\eqref{eq:ubaryiszero} implies that 
$$
n_{\varepsilon}(t,x,y) \xrightharpoonup[\varepsilon  \rightarrow 0]{} \rho(t,x) \, \delta_{\bar{y}(t,x)}(y) \quad \text{weakly in measures},
$$
where $\delta_{\bar{y}(t,x)}(y)$ is the Dirac delta centred at $y=\bar{y}(t,x)$.
\end{remark}

\paragraph{Relation between $\bar{y}(t,x)$ and $p(t,x)$.} Assumptions (2.3) ensure that ${\rm Supp}(p) \subseteq {\rm Supp}(\rho)$. Hence, evaluating~\eqref{eq:PDEu} at $y=\bar{y}(t,x)$ and using~\eqref{eq:ubaryiszero} and~\eqref{eq:uybaryiszero} we find 
\begin{equation}
\label{eq:Riszero}
R(\bar{y}(t,x),p(t,x)) = 0, \quad x \in {\rm Supp}(p).
\end{equation}
The monotonicity assumptions ensure that $p \mapsto R(\cdot,p)$ and $\bar{y} \mapsto R(\bar{y},\cdot)$ are both invertible. Therefore, relation \eqref{eq:Riszero} gives a one-to-one correspondence between $\bar{y}(t,x)$ and $p(t,x)$.

\paragraph{Transport equation for $\bar{y}$.}
Differentiating~\eqref{eq:PDEu} with respect to $y$, evaluating the resulting equation at $y=\bar{y}(t,x)$ and using~\eqref{eq:ubaryiszero} and~\eqref{eq:uybaryiszero} yields
\begin{equation}
\label{eq:PDEuatbary}
\partial^2_{yt} u(t,x,\bar{y}) - \hat{\mu}(\bar{y}) \, \partial_{x} p \, \partial^2_{yx} u(t,x,\bar{y}) = \partial_{y} R(\bar{y},p), \quad x\in {\rm Supp}(p).
\end{equation}
Moreover, differentiating~\eqref{eq:uybaryiszero} with respect to $t$ and $x$ we find, respectively,
\[
\partial^2_{ty} u(t,x,\bar{y}) + \partial^2_{yy} u(t,x,\bar{y}) \, \partial_{t} \bar{y}(t,x) = 0 \; \Rightarrow \; \partial^2_{yt} u(t,x,\bar{y}) = - \partial^2_{yy} u(t,x,\bar{y}) \, \partial_{t} \bar{y}(t,x)
\]
and
%\begin{equation}
%\label{eq:u2xy}
$$
\partial^2_{xy} u(t,x,\bar{y}) + \partial^2_{yy} u(t,x,\bar{y}) \, \partial_{x} \bar{y}(t,x) = 0 \; \Rightarrow \; \partial^2_{yx} u(t,x,\bar{y}) = - \partial^2_{yy} u(t,x,\bar{y}) \, \partial_{x} \bar{y}(t,x).
$$
%\end{equation}
Substituting the above expressions of $\partial^2_{yt} u(t,x,\bar{y})$ and $\partial^2_{yx} u(t,x,\bar{y})$ into~\eqref{eq:PDEuatbary} and using the fact that $\partial^2_{yy} u(t,x,\bar{y}) < 0$ gives the following transport equation for $\bar{y}(t,x)$ 
\begin{equation}
\label{eq:PDEbary}
\partial_{t} \bar{y} - \hat{\mu}(\bar{y}) \, \partial_{x} p \, \partial_{x} \bar{y} = \frac{1}{-\partial^2_{yy} u(t,x,\bar{y})} \partial_{y} R(\bar{y},p), \quad x \in {\rm Supp}(p).
\end{equation}

\paragraph{Travelling-wave problem.} Substituting the travelling-wave ansatz
\[
\rho(t,x) = \rho(z), \quad p(t,x) = p(z), \quad 
u(t,x,y) = u(z,y) \quad \text{and} \quad \bar{y}(t,x) = \bar{y}(z) \quad \text{with} \quad z = x - c \, t, \quad c>0
\]
into~\eqref{eq:PDEu}-\eqref{eq:Riszero} and~\eqref{eq:PDEbary} gives 
%\begin{equation}
%\label{eq:TWu}
$$
- \left(c + \hat{\mu}(y) p' \right) \partial_z u = R(y,p) + (\partial_y u)^2, \quad (z,y) \in \mathbb{R} \times (0,Y),
$$
%\end{equation}
%\begin{equation}
%\label{eq:ubaryiszeroTW}
$$
u(z,\bar{y}(z)) = \max_{y \in [0,Y]} u(z,y) = 0, \quad \partial_y u(z,\bar{y}(z)) = 0, \quad \partial_z u(z,\bar{y}(z)) = 0, \quad z \in {\rm Supp}{\left(\rho \right)}, 
$$
%\end{equation}
\begin{equation}
\label{eq:TWRiszero}
R(\bar{y}(z),p(z)) = 0, \quad z \in {\rm Supp}(p),
\end{equation}
\begin{equation}
\label{eq:TWbary}
- \left(c + \hat{\mu}(\bar{y}) p' \right) \bar{y}' = \frac{1}{-\partial^2_{yy} u(z,\bar{y})} \partial_{y} R(\bar{y},p), \quad z \in {\rm Supp}(p).
\end{equation}
We consider travelling-front solutions $\bar{y}(z)$ that satisfy~\eqref{eq:TWbary} subject to the following asymptotic condition
\begin{equation}
\label{eq:TWBCy}
\lim_{z \to - \infty} \bar{y}(z) =0,
%\bar{y}(-\infty) = arg\,max_{y \in [0,1]} R(y,\rho(-\infty)).
\end{equation}
so that, since $R(0,p_M)=0$, relation~\eqref{eq:TWRiszero} gives $\displaystyle{\lim_{z \to - \infty} p(z) = p_M}$.

\paragraph{Monotonicity of travelling-front solutions.} 
Differentiating~\eqref{eq:TWRiszero} with respect to $z$ gives 
\begin{equation}
\label{eq:TWRziszero}
\partial_y R(\bar{y}(z),p(z)) \bar{y}'(z) + \partial_{p} R(\bar{y}(z), p(z)) p'(z) = 0, \quad z \in {\rm Supp}(p).
\end{equation}
Substituting the expression of $p'$ given by~\eqref{eq:TWRziszero} into~\eqref{eq:TWbary} yields
\[
-c \ \bar{y}' + \hat{\mu}(\bar{y}) \ \dfrac{\partial_y R(\bar{y},p)}{\partial_{p} R(\bar{y},p)} \ \left(\bar{y}' \right)^2 = \frac{1}{-\partial^2_{yy} u(z,\bar{y})} \, \partial_{y} R(\bar{y},p),
\]
that is,
\begin{equation}
\label{eq:TWbary2}
\bar{y}' = \frac{-\partial_y R(\bar{y},p)}{c} \left(\frac{1}{-\partial^2_{yy} u(z,\bar{y})} + \dfrac{\hat{\mu}(\bar{y}) \left(\bar{y}' \right)^2}{-\partial_{p} R(\bar{y},p)} \right), \quad z \in {\rm Supp}(p).
\end{equation}
Since $\partial^2_{yy} u(z,\bar{y})<0$ and $\partial_y R(y,\cdot)<0$ for $y \in (0,Y]$, using~\eqref{eq:TWbary2} and the expression of $p'$ given by~\eqref{eq:TWRziszero} we find
\begin{equation}
\label{eq:TWbaryincrhodec}
\bar{y}'(z) > 0 \quad \text{and} \quad p'(z) < 0, \quad z \in {\rm Supp}(p).
\end{equation}

\paragraph{Position of the edge of the travelling front $p(z)$.} 
Relation~\eqref{eq:TWRiszero} and monotonicity results~\eqref{eq:TWbaryincrhodec} along with the fact that $R(Y,0)=0$ [{\it cf.} assumptions (2.7)] imply that the position of the edge of the travelling front $p(z)$ coincides with the unique point $\ell \in \mathbb{R}$ such that $\bar{y}(\ell)=Y$ and $\bar{y}(z) < Y$ on $(-\infty, \ell)$. Hence, ${\rm Supp}(p) = (-\infty, \ell)$. 

\paragraph{Minimal wave speed.} 
In the case where $R(y,p)$ is defined via (2.8), relation~\eqref{eq:TWRiszero} yields 
%\begin{equation}
%\label{eq:TWRiszerospec}
$$
p(z) = p_M \, r(\bar{y}(z)), \quad z \in {\rm Supp}(p).
$$
Therefore, ${\rm Supp}(p) = {\rm Supp}(r(\bar{y}))$.
%\end{equation}
Moreover, we have 
\[
\partial_{p} R(\cdot,p) = -\dfrac{1}{p_M} \quad \text{and} \quad \displaystyle{\partial_y R(\bar{y},\cdot) = \dfrac{{\rm d}}{{\rm d} y}} r(\bar{y}).
\]
Hence, recalling that $\hat{\mu}(y) := \dfrac{\mu(y)}{p_M}$, from equation \eqref{eq:TWbary2} we find
\begin{equation}
\label{eq:precex}
\mu(\bar{y}) \ \partial^2_{yy} u(z,\bar{y})\ \dfrac{{\rm d}}{{\rm d} y} r(\bar{y}) \ \left(\bar{y}'\right)^2 + c \, \partial^2_{yy} u(z,\bar{y}) \ {\bar{y}}' - \dfrac{{\rm d} }{{\rm d} y} r(\bar{y}) = 0, \quad z \in r(\bar{y}(z)).
\end{equation}
The following condition has to hold for the roots of~\eqref{eq:precex}, seen as an algebraic equation for $\bar{y}'(z)$, to be real
\[
c \geq 2 \ \left|\dfrac{{\rm d}}{{\rm d} y} r(\bar{y})\right| \sqrt{\dfrac{\mu(\bar{y})}{\left|\partial^2_{yy} u(z,\bar{y})\right|}}, \quad z \in r(\bar{y}(z)).
\]
This gives condition (4.4) on the wave speed.

\section{Methods used to solve numerically the non-local PDE (4.2)}
\label{appendix_PDE}
Adopting a time-splitting approach, which is based on the idea of decomposing the original problem into simpler subproblems that are then sequentially solved at each time-step, we decompose the non-local PDE (4.2) posed on $\Omega := (0,T]\times(0, X)\times(0,Y)$, with $T=8$, $X=25$ and $Y=1$, into two parts -- {\it i.e.} the diffusion-advection part corresponding to the following non-local PDE
\begin{equation}\label{dyn:convection}
\begin{cases}
\partial_t n_\varepsilon - \hat{\mu}(y) \, \partial_x(n_\varepsilon \, \partial_x p_\varepsilon)=\varepsilon \, \partial_{yy}^2 n_\varepsilon,
\\
p_\varepsilon = \Pi(\rho_\varepsilon), \quad \rho_\varepsilon(t,x) = \int_0^Y \, n_\varepsilon(t,x,y)\, dy.
\end{cases}
\end{equation}
and the reaction part corresponding to the following integro-differential equation
\begin{equation}\label{dyn:grow}
\begin{cases}
\varepsilon \, \partial_t n_\varepsilon = R(y,p_\varepsilon) \, n_\varepsilon,
\\
p_\varepsilon = \Pi(\rho_\varepsilon), \quad \rho_\varepsilon(t,x) = \int_0^Y \, n_\varepsilon(t,x,y)\, dy.
\end{cases}
\end{equation}
We complement~\eqref{dyn:convection} with zero Neumann boundary conditions at $x=0$ (we expect a constant step), $y=0$ and $y=Y$. 
With the ansatz $n_\varepsilon(t,x,y) = e^{\frac{u_\varepsilon(t,x,y)}{\varepsilon}}$, the integro-differential equation~\eqref{dyn:grow} can be rewritten in the following alternative form
\begin{equation}\label{dyn:grow_u}
\begin{cases}
\partial_t u_\varepsilon = R(y, p_\varepsilon),
\\
p_\varepsilon = \Pi(\rho_\varepsilon), \quad \rho_\varepsilon(t,x) = \int_0^Y \, e^{\frac{u_\varepsilon(t,x,y)}{\varepsilon}}\, dy.
\end{cases}
\end{equation}

\paragraph{Preliminaries and notation} We denote by $\llbracket k_1, k_2\rrbracket$ the set of integers between $k_1$ and $k_2$. We discretise $\Omega$ via a uniform structured grid of steps $\Delta{t}$, $\Delta{x}$, $\Delta{y}$ whereby $t_h=h\Delta{t}$ and the $(j, k)$-th cell is
$$
K_{j,k} = (x_{j-1},x_{j})\times (y_{k-1},y_{k})\quad \text{with}\quad x_j = j\Delta{x}, \quad y_k = k \Delta{y}, 
$$
where $j \in \llbracket1, m_x\rrbracket$ and $k\in \llbracket1, m_y\rrbracket$, 
$\Delta{x} = \frac{X}{m_x}$, $\Delta{y} = \frac{Y}{m_y}$ and $m_x,m_y\in \mathbb{N}$. In particular, given $\Omega := (0,T]\times(0, X)\times(0,Y)$, with $T=8$, $X=25$ and $Y=1$, we choose $\Delta{t}=10^{-4}$, $\Delta{x}=0.01$ and $\Delta{y}=0.02$. Moreover, we let $N_{\varepsilon j,k}^h$ be the numerical approximation of the average of $n_\varepsilon(t_h, x, y)$ over the cell $K_{j,k}$ and 
\[
\rho_{\varepsilon j}^{h} = \Delta{y} \sum_{k=1}^{m_y} N_{\varepsilon j,k}^{h}.
\]
be the average of $\rho_\varepsilon(t_h, x)$ over the interval $(x_{j-1},x_{j})$. For simplicity of notation, in the remainder of this section we drop the subscript $\varepsilon$.

\paragraph{Numerical scheme} \ \\
{\bf Step 1} We first solve numerically~\eqref{dyn:convection} by using the following implicit-explicit scheme
\begin{equation}\label{scheme:conv}
 \frac{N_{j,k}^{*} - N_{j,k}^h}{\Delta{t}} - \hat{\mu}_k\frac{ \delta_x P_{j+\frac12}^{h} N_{j+\frac12, k}^{*} - \delta_x P_{j-\frac12}^{h} N_{j-\frac12, k}^{*} }{\Delta{x}} 
	=
	\varepsilon \frac{N_{j,k+1}^{*} - 2 N_{j,k}^{*} + N_{j,k-1}^{*}}{(\Delta{y})^2}.
\end{equation}
where $\hat{\mu}_k = \hat{\mu}(y_{k})$, $\delta_x P_{j+\frac12}^{h}=(P_{j+1}^h - P_{j}^h)/\Delta{x}$ and 
\begin{equation*}
 N_{j+\frac12, k}^* = 
 \begin{cases}
 N_{j, k}^*, &\text{ if } \delta_x P_{j+\frac12}^{h} \le0, \\
 N_{j+1, k}^*, &\text{ if } \delta_x P_{j+\frac12}^{h} >0.
 \end{cases}
\end{equation*}
Zero-flux/Neumann boundary conditions are implemented at $x=0$, $y=0$ and $y=Y$.

{\bf Step 2} 
Starting from $U_{j,k}^* = \varepsilon \ln \left(N_{j,k}^{*} \right)$, where $N_{j,k}^{*}$ is obtained via~\eqref{scheme:conv}, 
we solve numerically~\eqref{dyn:grow_u} using the following implicit scheme 
\begin{equation}\label{scheme:rho_growth} 
\begin{cases}
&U_{j,k}^{n+1} = U_{j,k}^n + \Delta{t} \ 
R(y_{k-\frac12}, P_j^{n+1}), \\
& P_j^{h+1} = \Pi(\rho_j^{h+1}), 
\quad 
\rho_j^{h+1} = \Delta{y} \sum_{k=1}^{m_y} e^{\frac{U_{j,k}^{h+1}}{\varepsilon}}.
\end{cases}
\end{equation}
Substituting the first equation in \eqref{scheme:rho_growth} into the second equation yields 
$$
 \rho_{j}^{h+1} = \Delta{y} \sum_{k=1}^{m_y} \exp\left( \dfrac{U_{j,k}^{*}+\Delta{t} \ R\left(y_{k}, P_{j}^{h+1}\right)}{\varepsilon}\right),
$$
from which $\rho_{j}^{h+1}$ and $P_{j}^{h+1}$ are computed. Straightforward calculations then lead to $U_{j,k}^{n+1}$ and $N_{j,k}^{n+1}$.

\end{document}